\documentclass[a4paper, 11pt]{article}
\usepackage{jheppub}% for details on the use of the package, please
\usepackage[T1]{fontenc} % if needed
\usepackage{graphicx}
\usepackage{tikz}
\usepackage{amsmath,amssymb,multirow,xspace,slashed,array,booktabs}
\usepackage[colorlinks=true,urlcolor=blue,anchorcolor=blue,citecolor=blue,filecolor=blue,linkcolor=blue,menucolor=blue,pagecolor=blue]{hyperref}
\usepackage{natbib}
\setcitestyle{square, comma, numbers,sort&compress, super}
\usepackage{booktabs}
\usepackage{blindtext, rotating}
\usepackage{appendix}
\usepackage{afterpage}
\usepackage{enumitem}
\usepackage{marvosym}
\usepackage{bbold}
\usepackage{slashed}
\usepackage{physics}
\usepackage{verbatim}
\usepackage{indentfirst}
\usepackage{bm}
\usepackage{soul} %scratch out text
\usepackage[normalem]{ulem}
\usepackage{pifont}% http://ctan.org/pkg/pifont
\usepackage{cancel}
\usepackage{makecell}
\usepackage[capitalise]{cleveref}
\crefname{equation}{Eq.}{Eqs.} % capitalize "E", no period
\crefrangelabelformat{equation}{(#3#1#4--#5#2#6)}

\newcommand{\gev}{\mathrm{GeV}}

\newcommand{\fig}[2]{Fig.~\hyperref[fig:#1]{\ref*{fig:#1}(#2)}}
\newcommand{\SC}{\mathbb{S}}

\title{\boldmath 
\Large {The Discriminant Power of Bubble Wall Velocities:}\\
{\Large Gravitational Waves and Electroweak Baryogenesis}
}

\abstract{
A precise determination of the bubble wall velocity $v_w$ is crucial for making accurate predictions of the baryon asymmetry and gravitational wave (GW) signals in models of electroweak baryogenesis (EWBG).
Working in the local thermal equilibrium approximation, we exploit entropy conservation to present efficient algorithms for computing $v_w$, significantly streamlining the calculation.
We then explore the parameter dependencies of $v_w$, focusing on two sample models capable of enabling a strong first-order electroweak phase transition: a $\mathbb{Z}_2$-symmetric singlet extension of the SM, and a model for baryogenesis with CP violation in the dark sector. 
We study correlations among $v_w$ and the two common measures of phase transition strength, $\alpha_n$ and $v_n/T_n$.
Interestingly, we find a relatively model-insensitive relationship between $v_n/T_n$ and $\alpha_n$. 
We also observe an upper bound on $\alpha_n$ for the deflagration/hybrid wall profiles naturally compatible with EWBG, the exact value for which varies between models, significantly impacting the strength of the GW signals.
In summary, our work provides a framework for exploring the feasibility of EWBG models in light of future GW signals.
}

\author[a,b,c,d,e]{Marcela Carena}
\author[c,d,f]{Aurora Ireland}
\author[c,d]{Tong Ou}
\author[b]{Isaac R. Wang}
\affiliation[a]{Perimeter Institute for Theoretical Physics, Waterloo, Ontario N2L 2Y5, Canada}
\affiliation[b]{Fermi National Accelerator Laboratory,  P. O. Box 500, Batavia, IL 60510, USA}
\affiliation[c]{Department of Physics, University of Chicago, Chicago, IL 60637, USA}
\affiliation[d]{Enrico Fermi Institute, University of Chicago, Chicago, IL 60637, USA}
\affiliation[e]{Kavli Institute for Cosmological Physics, University of Chicago, Chicago, IL 60637}
\affiliation[f]{Stanford Institute for Theoretical Physics, Stanford University, Stanford, CA 94305}
\emailAdd{mcarena@perimeterinstitute.ca}
\emailAdd{anireland@stanford.edu}
\emailAdd{tongou@uchicago.edu}
\emailAdd{isaacw@fnal.gov}
\preprint{FERMILAB-PUB-25-0266-T}
\begin{document}
\maketitle
                                                                                                                                                                                                                                                              
\flushbottom

%%%%%%%%%%%%%%%%%%%%%%%%%%%%%%%%%%%%%%%%%%%%%%%%%%%%%%%%%
%%%%%%%%%%%%%%%%%%%%%%%%%%%%%%%%%%%%%%%%%%%%%%%%%%%%%%%%%
\section{Introduction}

The Standard Model (SM) of particle physics provides an excellent description of the particles constituting ordinary matter and their interactions, but fails to explain how this matter came to be. That is, the SM does not provide a complete framework for baryogenesis --- the physical process responsible for generating the observed baryon asymmetry of the universe (BAU). A successful baryogenesis requires baryon number violation, charge (C) and charge-parity (CP) violation, and a departure from thermal equilibrium~\cite{Sakharov:1967dj}. In principle, the electroweak phase transition (EWPT) provides an ideal scenario in which all three of these Sakharov conditions can be satisfied, and electroweak baryogenesis (EWBG) remains among the most attractive ways of accounting for the BAU~\cite{Kuzmin:1985mm,Shaposhnikov:1986jp,Farrar:1993hn,Farrar:1993sp}.

The CP-violating mixing angles and Yukawa couplings in the SM quark sector are incapable of triggering a sufficiently large CP asymmetry to source EWBG, however~\cite{Gavela:1993ts,Huet:1994jb,Gavela:1994ds,Gavela:1994dt,Gavela:1994yf}. Furthermore, presuming only the SM particle content, the EWPT is a smooth crossover rather than a strong first-order transition needed to provide out-of-equilibrium conditions~\cite{Kajantie:1993ag,Farakos:1994kj,Jansen:1995yg,Kajantie:1995kf,Rummukainen:1996sx,Kajantie:1996mn,Gurtler:1997hr,Csikor:1998eu,Laine:1998vn,Laine:1998qk,Rummukainen:1998as,Fodor:1999at}. While EWBG cannot be realized in the SM, it can easily be accommodated in beyond the Standard Model (BSM) extensions which provide additional sources of CP violation and additional particle content to make the transition strongly first order. See Refs.~\cite{Anderson:1991zb,Pietroni:1992in,Espinosa:1993bs,McDonald:1993ey,Choi:1993cv,Joyce:1994bi,Cline:1995dg,Basler:2016obg,Basler:2017uxn,Jeong:2018jqe,Jeong:2018ucz,Grzadkowski:2018nbc,Basler:2019iuu,Carena:2019une,Chao:2019smr,Xie:2020wzn,Xie:2020bkl,Fernandez-Martinez:2020szk,Basler:2021kgq,Biermann:2022meg,Anisha:2022hgv,Anisha:2023vvu,Fernandez-Martinez:2022stj,Harigaya:2023bmp} for sample BSM extensions which result in a successful EWBG and Refs.~\cite{Shu:2006mm,Long:2017rdo,Baldes:2018nel,Glioti:2018roy,Hall:2019ank,Hall:2019rld,Fujikura:2021abj,Matsedonskyi:2022btb,Harigaya:2022wzt,Bhandari:2025ufp} for models with EW-like first-order phase transitions triggering baryogenesis.

In addition to potentially accommodating baryogenesis, a strong first-order EWPT may also result in observable gravitational wave (GW) signals. First-order phase transitions proceed via the nucleation of true vacuum bubbles which collide and merge, allowing the gradient energy in the scalar field to source GWs~\cite{Kamionkowski:1994,Caprini:2007xq,Huber:2008hg}. During the first-order phase transition, expanding sound shells of fluid kinetic energy also propagate and collide, providing an additional source of GWs~\cite{Hindmarsh:2013xza,Hindmarsh:2016lnk}. This acoustic stage may produce shocks and turbulence in the plasma, which in turn also produce GWs~\cite{Kosowsky:2002,Caprini:2006,Caprini:2009yp}. The relative contribution from each of these sources depends on the properties of the phase transition. GW production in phase transitions with significant supercooling is typically dominated by the bubble collision stage, while for thermal vacuum transitions in the plasma, acoustic effects provide the dominant contribution.

Both the final baryon asymmetry and the size of the GW signal depend crucially on the bubble wall velocity $v_w$.
Significant GW production is only possible for sufficiently fast-moving bubble walls. Meanwhile, smaller $v_w$ values are typically associated with larger baryon asymmetries, since slower walls allow for more efficient diffusion of particle asymmetries.
This has led to the standard lore that there exists a tension between EWBG and observable GW signals \cite{Huet:1995sh}, though the extent of this tension remains the subject of active debate~\cite{Cline:2020jre,Li:2024mts}.
In certain contexts, a larger $v_w$ has been found to correlate positively with a larger baryon asymmetry~\cite{Cline:2021iff, Fromme:2006wx}\footnote{Faster wall velocities increase particle scattering rates near the wall, potentially enhancing the CP asymmetry.
    Additionally, a larger $v_w$ can lead to thinner walls, resulting in an enhanced baryon asymmetry.}.
At any rate, a precise determination of $v_w$ remains the key task for accurate predictions of the final baryon asymmetry and the GW signal.

The wall velocity should, in theory, be calculable from first principles. Equations of motion (EOM) for the wall-fluid system follow from conservation laws and have been known for almost 30 years~\cite{Ignatius:1993qn,Moore:1995si}.
These equations were applied to study the hydrodynamics of bubble growth in the pioneering work of Ref.~\cite{Espinosa:2010hh}, which developed a unified framework classifying fluid profiles into deflagration, hybrid, and detonation solutions, laying the foundation for modern computations of the wall velocity.
In recent years, a first-principles framework for bubble wall dynamics has been developed and applied in the context of a singlet scalar extension with $\mathbb{Z}_2$ symmetry~\cite{Laurent:2022jrs}.
This framework is broadly applicable so long as the plasma's deviation from equilibrium can be treated perturbatively.
It was further confirmed in this work that out-of-equilibrium contributions to the hydrodynamic obstruction are subdominant\footnote{For recent progress regarding out-of-equilibrium contributions, see Ref.~\cite{DeCurtis:2022hlx,DeCurtis:2023hil,DeCurtis:2024hvh}}, justifying the treatments working in local thermal equilibrium (LTE).
This is fortuitous, since in LTE entropy conservation provides an additional matching condition that greatly simplifies the computation of the bubble wall velocity~\cite{Ai:2021kak,Ai:2023see,Ai:2024shx}.

In this work, we present an in-depth exploration of the parameter dependencies of the bubble wall velocity and its implications for baryogenesis and the GW signal.
We begin in Sec.~\ref{sec:bubblewallreview} by reviewing the hydrodynamics of an expanding bubble wall.
We describe how to calculate the terminal wall velocity $v_w$ based on matching conditions following from energy-momentum conservation and the scalar equation of motion --- or equivalently from entropy conservation when working in the LTE approximation  --- presenting efficient algorithms to calculate $v_w$ for each fluid profile class.
In Sec.~\ref{sec:vw computation} we demonstrate concretely how to use these algorithms by solving for $v_w$ in two specific models.
We examine first a $\mathbb{Z}_2$-symmetric real singlet extension of the SM, which is the minimal scenario capable of rendering the EWPT first order.
By comparing with known results in the literature, we verify explicitly the validity of the LTE approximation and our algorithm utilizing entropy conservation.
We then calculate $v_w$ in the model of Refs.~\cite{Carena:2018cjh,Carena:2019xrr,Carena:2022qpf}, which introduces CP violation in a dark sector to realize baryogenesis whilst evading the strong constraints on electric dipole moments.
We will find that the typical wall velocities are much higher than the commonly assumed values for EWBG, and so in Sec.~$\ref{sec:BAU}$ we examine the implications for baryogenesis in this model.
We compute the BAU in different computational frameworks and comment on their applicability depending on the value of $v_w$.
Finally in Sec.~\ref{sec:GWsignal} we examine the consequences for the GW signal.
We conclude in Sec.~\ref{sec:conclusion} with further discussion of the results and comments on future directions.

%%%%%%%%%%%%%%%%%%%%%%%%%%%%%%%%%%%%%%%%%%%%%%%%%%%%%%%%%
%%%%%%%%%%%%%%%%%%%%%%%%%%%%%%%%%%%%%%%%%%%%%%%%%%%%%%%%%
\section{Bubble Wall Dynamics}\label{sec:bubblewallreview}

Following bubble nucleation, the potential energy differential drives the bubble wall to accelerate into the unbroken phase. As it does so, it perturbs and heats the surrounding plasma, which in turn exerts a frictional force on the wall. Once this frictional force balances the driving force, the bubble wall reaches a terminal velocity $v_w$.

The terminal bubble wall velocity $v_w$ can be computed from first principles.
In the bubble wall frame, one defines the fluid velocity near the bubble wall both in the symmetric and broken phases, $v_+$ and $v_-$, respectively, and relates these quantities through matching conditions.
We then relate the fluid velocities with the instantaneous bubble wall velocity depending on the type of plasma motion. The matching conditions are derived by means of energy-momentum conservation, considering the plasma and scalar field as a closed system.
The scalar field equation of motion (EOM) also provides an independent matching condition.
As we will explicitly show, for a system in local thermal equilibrium (LTE), one can equivalently use entropy conservation as a matching condition in place of EOM. This approach provides a much more efficient way of performing numerical calculations.

%%%%%%%%%%%%%%%%%%%%%%%%%%%%%%%%%%%%%%%%%%%%%%%%%%%%%%%%%
\subsection{Energy-Momentum Conservation}\label{sec:EMconservation}

Energy-momentum conservation in the closed scalar-fluid system is expressed through the following conservation law,
\begin{equation}\label{eq:conservation}
    \partial_\mu T^{\mu \nu} = \partial_\mu\left(T_\phi^{\mu \nu} + T_f^{\mu \nu}\right) = 0 \,,
\end{equation}
where in the second term we have decomposed the energy-momentum tensor (EMT) into scalar $\phi$ and fluid $f$ contributions. The scalar field EMT with zero-temperature potential $V_0(\phi)$\footnote{$V_0$ includes both the tree-level potential $V_{\rm tree}$ and quantum corrections, most notably the 1-loop Coleman Weinberg potential $V_{\rm CW}$.} is given by\footnote{We presume the phase transition proceeds sufficiently quickly relative to the Hubble rate that the spacetime is approximately Minkowski $\eta^{\mu \nu}$. We use the mostly-negative signature $(+,-,-,-)$ to match the convention in the literature.}
\begin{equation}\label{eq:Tphi}
    T_\phi^{\mu \nu} = \partial^\mu \phi \partial^\nu \phi - \eta^{\mu \nu} \left( \frac{1}{2} (\partial \phi)^2 - V_0\right) \,,
\end{equation}
while the fluid contribution is
\begin{equation}\label{eq:Tf}
    T_f^{\mu \nu} = \sum_i n_i \int \frac{d^3k}{(2\pi)^3} \frac{k^\mu k^\nu}{E_{i,k}} f_i(k,x) \,,
\end{equation}
where the sum runs over all dynamical particle species in the plasma, $n_i$ is the particle degrees of freedom for species $i$, and $f_i(k,x)$ is the phase space distribution function for species $i$. \cref{eq:Tf} is the most general expression for the fluid EMT and can be used for general out-of-equilibrium calculations of bubble wall dynamics, which we review in \cref{app:non_LTE}.

In what follows, we will assume the fluid is in local thermal equilibrium (LTE), i.e. $f_i(k,x)=f_i^{\rm eq}(k,x)$, with equilibrium distribution
\begin{equation}
    f_i^{\rm eq}(k,x) = \frac{1}{e^{E_i/T}\mp 1} \,,
\end{equation}
where the $-$ ($+$) sign is for bosons (fermions) and $E_i = \sqrt{\vec{k}^2 + m_i^2}$. This approximation is justified for sub-luminal wall velocities $v_w < 1$, for which equilibrium effects give the dominant contribution to the net pressure controlling bubble expansion. As shown in Ref.~\cite{Laurent:2022jrs}, as well as in this work, the LTE approximation can accurately reproduce the full out-of-equilibrium result for the terminal wall velocity in this regime. Although out-of-equilibrium contributions may become important for ultra-relativistic bubbles with $v_w\sim 1$, this scenario is beyond the scope of this work\footnote{It was previously thought that the frictive force from the thermal plasma, important for preventing runaway bubble walls, arose only when there was a departure from thermal equilibrium in the plasma. This is now known not to be the case \cite{Balaji:2020,Ai:2021kak}.}.

In LTE, we can equivalently formulate the fluid EMT of \cref{eq:Tf} in terms of local thermodynamic quantities. To do so, we first identify the fluid pressure as the (negative of the) fluid free energy $\mathcal{F}_f$, which coincides with the finite temperature contribution to the effective potential for the scalar field $V_T$, as discussed in \cref{app:freeenergy},
\begin{equation}
    \label{eq:pressure_free_energy}
    p_f = - \mathcal{F}_f = - V_T \,.
\end{equation}
The fluid enthalpy density $w_f$, entropy density $s_f$, and energy density $e_f$ are defined in terms of the pressure as
\begin{equation}\label{eq:thermquantities}
    w_f = T \frac{\partial p_f}{\partial T} \,, \,\,\,\,\,\, s_f = \frac{\partial p_f}{\partial T} \,, \,\,\,\,\,\, e_f = T \frac{\partial p_f}{\partial T} -p_f \,,
\end{equation}
such that $w_f = e_f + p_f$. It is also useful to note that the energy density and pressure for the scalar field are $e_\phi = V_0(\phi)$ and $p_\phi = - V_0(\phi)$, satisfying the vacuum equation of state. Hence the total pressure $p = p_f + p_\phi$ and energy density $e = e_f + e_\phi$ are,
\begin{equation}\label{eq:totalthermquantities}
    p = p_f - V_0 \,, \,\,\,\, e = e_f + V_0 \,.
\end{equation}
The enthalpy and entropy densities receive contributions only from the fluid, such that $w = w_f$ and $s = s_f$. Using these quantities, the fluid EMT in LTE can be written in a perfect fluid form
\begin{equation}\label{eq:perfectfluid}
    T_{f,\text{LTE}}^{\mu \nu} = w_f \, u^\mu u^\nu - p_f \, \eta^{\mu \nu} \,,
\end{equation}
with $u_\mu$ the fluid 4-velocity, which is normalized as $u^\mu u_\mu = 1$.
The divergence entering the EMT conservation law is then
\begin{equation}
    \partial_\mu T_{f,\text{LTE}}^{\mu \nu} = \partial_\mu (w_f u^\mu u^\nu) - \partial^\nu \phi \frac{\partial p_f}{\partial \phi} - s_f \partial^\nu T \,,\label{eq:partial_Tf_LTE}
\end{equation}
where we have allowed for the temperature to have a non-trivial profile in spacetime and used the definition of the entropy density above.
Combining this result with the scalar field contribution,
\begin{equation}\label{eq:divTphi}
    \partial_\mu T^{\mu \nu}_\phi = \partial^\nu \phi \left( \Box \phi + \frac{\partial V_0}{\partial \phi} \right) \,,
\end{equation}
the statement of EMT conservation becomes
\begin{equation}\label{eq:startproof}
    \partial^\nu \phi \left( \Box \phi + \frac{\partial V_{\rm eff}}{\partial \phi} \right) + \partial_\mu (w u^\mu u^\nu) - s \partial^\nu T = 0 \,,
\end{equation}
where we have used the fact that $p_f = - V_T$ and combined $V_0 + V_T = V_{\rm eff}$. We will return to this expression shortly when we turn to derive the condition for entropy conservation.

%%%%%%%%%%%%%%%%%%%%%%%%%%%%%%%%%%%%%%%%%%%%%%%%%%%%%%%%%
\subsection{Matching Conditions in LTE}\label{sec:matching_conds}
In this subsection, we focus on the steady state where the bubble wall expands at constant velocity, and derive the matching conditions for the fluid velocities ($v_{\pm}$, $+/-$ for symmetric/broken phase) and temperatures ($T_{\pm}$) on each side of the bubble wall under the LTE approximation. There are two matching conditions coming from the EMT conservation, and a third one from the combination of the scalar field EOM and EMT conservation. The third condition is also identified as the entropy conservation condition, which will be shown to be equivalent to the scalar field EOM in LTE.

We choose to work in the wall rest frame, which is an inertial frame for a steady-state bubble\footnote{See \cref{app:frames} for a general review of the reference frames for the scalar-fluid system.}. This reference frame has the benefit that all quantities are time-independent and depend only on the coordinate corresponding to the direction of bubble expansion, which is taken to be the $z$-direction. Thus, in the wall frame, EMT conservation \cref{eq:conservation} leads to just two independent equations
\begin{equation}\label{eq:partialz_T}
    \partial_z T^{z0}=0 \,, \quad \partial_z T^{zz}=0 \,.
\end{equation}
We also note that in the wall frame, the fluid 4-velocity $u^\mu$ can be parameterized as
\begin{align}\label{eq:umu}
    u^\mu(z)=\gamma(z)\left(1,0,0,-v(z)\right),
\end{align}
with $v$ the physical velocity in the $z$-direction and $\gamma=1/\sqrt{1-v^2}$ the Lorentz factor, so that it satisfies the normalization condition $u^\mu u_\mu=1$.

\cref{eq:partialz_T} leads to two matching conditions as follows. Expanding the scalar and fluid contributions explicitly using Eqs.~(\ref{eq:Tphi}) and (\ref{eq:perfectfluid}), we have\footnote{Note that the thermodynamic quantities appearing here correspond to the total scalar-fluid system and are defined in \cref{eq:totalthermquantities}.}
\begin{subequations}
    \begin{equation}
        \partial_z \left( w \gamma^2 v \right) = 0 \,,
    \end{equation}
    \begin{equation}
        \partial_z \left( \frac{1}{2} (\partial_z \phi)^2 + w \gamma^2 v^2 + p \right) = 0 \,.
    \end{equation}
\end{subequations}
Integrating over $z$, we obtain the following matching conditions\footnote{In deriving \cref{eq:matchcond2}, we have used the fact that $\partial_z \phi = 0$ away from the wall.}
\begin{subequations}\label{eq:matchcond23_w}
    \begin{equation}\label{eq:matchcond1}
        \boxed{\text{Condition 1:}\quad w_+ \gamma_+^2 v_+ = w_- \gamma_-^2 v_- }\,,
    \end{equation}
    \begin{equation}\label{eq:matchcond2}
        \boxed{\text{Condition 2:}\quad  w_+ \gamma_+^2 v_+^2 + p_+ = w_- \gamma_-^2 v_-^2 + p_-} \,,
    \end{equation}
\end{subequations}
where we denote quantities in the symmetric/broken phase with $+/-$. This system of equations can be equivalently formulated as \cite{Espinosa:2010hh}
\begin{subequations}\label{eq:matchcond23_v}
    \begin{equation}\label{eq:mc1}
        v_+ v_- = \frac{p_+ - p_-}{e_+ - e_-} \,,
    \end{equation}
    \begin{equation}\label{eq:mc2}
        \frac{v_+}{v_-} = \frac{e_- + p_+}{e_+ + p_-} \,.
    \end{equation}
\end{subequations}

A third matching condition can be derived by imposing scalar field EOM on top of EMT conservation.
Returning to the condition of EMT conservation in \cref{eq:startproof}, we multiply both sides by $u_\nu$. From the normalization condition $u_\mu u^\mu = 1$, it follows that $u_\nu \partial_\mu u^\nu = 0$, and so one can show that
\begin{equation}\label{eq:entropyproof}
    u_\nu \partial^\nu \phi \left( \Box \phi + \frac{\partial V_{\rm eff}}{\partial \phi} \right) + T \partial_\mu S^\mu = 0 \,,
\end{equation}
where we have defined the entropy current $S^\mu = u^\mu s$. The quantity in parenthesis can be identified as the scalar field EOM in LTE, which should vanish independently: $\Box \phi + \partial_\phi V_{\rm eff} =0$, leaving us the conservation condition for the entropy
\begin{equation}\label{eq:entropyconservation}
    \partial_\mu S^\mu = 0 \,.
\end{equation}
This demonstrates
the equivalence of the scalar EOM with entropy conservation in LTE under EMT conservation, which was first identified in Ref.~\cite{Ai:2021kak} as a useful matching condition for calculating the bubble wall velocity. Entropy conservation is also generally expected for systems in thermal equilibrium based on the second law of thermodynamics. In the wall frame, Eq.~(\ref{eq:entropyconservation}) reads
\begin{equation}\label{eq:entropycons}
    \partial_z \left( s \gamma v \right) = 0 \,,
\end{equation}
which upon integrating over the wall becomes
\begin{equation}\label{eq:matchcond3}
    \boxed{\text{Condition 3:}\quad s_+ \gamma_+ v_+ = s_- \gamma_- v_-} \,.
\end{equation}
Using $s = w/T$ and Eq.~(\ref{eq:matchcond1}), this condition can equivalently be written as
\begin{equation}\label{eq:mc3}
    \gamma_+ T_+ = \gamma_- T_- \,.
\end{equation}

%%%%%%%%%%%%%%%%%%%%%%%%%%%%%%%%%%%%%%%%%%%%%%%%%%%%%%%%%
\subsection{Equation of State}\label{sec:EOS}

It would be useful to connect the matching conditions to the strength and temperature of the phase transition. For this we must specify the equations of state (EOS) for the fluid and the scalar field that express the thermodynamic quantities in terms of the temperature and other variables entering the effective potential.

We start from the simplest ``bag'' approximation where the fluid and scalar contributions are separable. In this case, the fluid can be well modeled as a relativistic gas, with pressure and energy given by
\begin{align}
    p^f_{\pm}=\frac{1}{3}a_{\pm}T_{\pm}^4,\quad e^f_{\pm}=a_{\pm}T_{\pm}^4.
\end{align}
In the bag model, all the species lighter than the temperature are counted into the relativistic degrees of freedom $a_{\pm}$, while the species heavier than the temperature are considered to be Boltzmann suppressed and have negligible contributions.
Since the fluid pressure is directly related to the free energy as given by \cref{eq:pressure_free_energy}, $a_{\pm}$ is given by
\begin{equation}\label{eq:baga}
    a_\pm = \frac{\pi^2}{30} \left( \sum_{B \in \pm} n_B + \frac{7}{8} \sum_{F \in \pm} n_F \right) \,,
\end{equation}
such that $(-p_{\pm}^f)$ is equal to the zeroth-order term of the high-temperature expansion of the free energy given by \cref{eq:free_energy_FB}.
Meanwhile, the scalar field energy is denoted as $e_{\pm}^\phi=\epsilon_{\pm}$, and its pressure satisfies the vacuum EOS: $p_{\pm}^\phi=-e_{\pm}^\phi=-\epsilon_{\pm}$. In the bag model, the scalar field contribution is independent of the fluid, and thus is determined by the zero-temperature scalar potential $V_0(\phi)$, which is conventionally normalized such that the scalar field energy in the broken phase vanishes, $\epsilon_-=V_0(\phi_-)=0$.
Summing the fluid and scalar field contributions, we get the EOS for the total pressure and energy of the scalar-fluid system as
\begin{equation}\label{eq:EOSgeneral}
    p_\pm = \frac{1}{3} a_\pm T_\pm^4 - \epsilon_\pm \,, \,\,\,\, e_\pm = a_\pm T_\pm^4 + \epsilon_\pm \,.
\end{equation}

Comparing the bag model expression for $p_\pm = - V_{\rm eff}(\phi_{\pm},T_{\pm})$ in Eq.~\eqref{eq:EOSgeneral} to the general form of $V_{\rm eff}$ in Eq.~(\ref{eq:Veffgenform}), one can easily see that the bag EOS neglects contributions from finite particle masses. This is problematic when there exist particles in the plasma with masses comparable to the temperature $m \sim T$, as is the case for many realistic particle models. In particular, there exist several such particles in the SM EWPT, for example the massive gauge bosons $W^{\pm}$ and $Z^0$. Systems with such intermediate-mass particles can still be described by the EOS given by \cref{eq:EOSgeneral}, but with generalized temperature-dependent definitions of $a_{\pm}$ and $\epsilon_{\pm}$ \cite{Espinosa:2010hh}:
\begin{equation}
    \label{eq:general_def_a_ep}
    a_\pm \equiv \frac{3}{4 T_\pm^3} \frac{\partial p_\pm}{\partial T_\pm} = \frac{3 w_\pm}{4 T_\pm^4} \,, \,\,\,\, \epsilon_\pm \equiv \frac{1}{4} (e_\pm - 3 p_\pm) \,.
\end{equation}
We note here that $a_\pm$ still serves as the coefficient of $T^4$ by this definition.
In this general definition, however, $a_\pm$ depends on field value and temperature.
$\epsilon_\pm$ is now defined to contain all the remaining things in $p_\pm$ other than $a_\pm T^4/3$.
The physical meaning of $\epsilon_\pm$ is apparent by identifying it with the trace of stress-energy tensor, $\epsilon_\pm = \eta_{\mu \nu} T^{\mu \nu}(\phi_\pm, T_\pm)$.
One can easily check that Eq.~\eqref{eq:general_def_a_ep} reduces to the bag model once only zeroth-order terms of $T$ in $V_{\rm eff}$ is kept.

Now we apply the EOS to the matching conditions derived in the previous section. For convenience, we introduce the following two parameters
\begin{equation}
    r = \frac{w_+}{w_-} \,,
\end{equation}
\begin{equation}\label{eq:generalalpha}
    \alpha_+ = \frac{4(\epsilon_{+}-\epsilon_{-})}{3w_+} \,.
\end{equation}
where $r$ represents the enthalpy ratio between the symmetric and broken phases.
The $\alpha_+$ is defined as the difference of the trace of stress-energy tensor normalized by $3/4$ of the enthalpy, and is widely used in gravitational wave calculation.
$\alpha_+$ reduces to the released vacuum energy density normalized by the radiation energy in the bag model.
$\alpha_+$ is commonly used to characterize the phase transition strength, especially in the GW calculations. In the following sections, we will study its correlation with another phase transition strength parameter commonly used in the EWBG studies, $v_n/T_n$, which characterizes the degree to which the electroweak sphaleron process is suppressed in the broken phase.

Substituting the general EOS into the two matching conditions of \cref{eq:matchcond23_v} and using these new parameters yields the relations
\begin{subequations}
    \begin{equation}\label{eq:bagEOSmatching1}
        v_+ v_- = \frac{1 - (1 - 3 \alpha_+) r}{3 - 3(1+\alpha_+) r} \,,
    \end{equation}
    \begin{equation}\label{eq:bagEOSmatching2}
        \frac{v_+}{v_-} = \frac{3 + (1 - 3 \alpha_+)r}{1 + 3(1+ \alpha_+) r} \,.
    \end{equation}
\end{subequations}
These can be combined by eliminating $r$ to give \cite{Espinosa:2010hh}
\begin{equation}\label{eq:vplus}
    v_+ = \frac{1}{6 (1 + \alpha_+) v_-} \left( 1 + 3 v_-^2 \pm \sqrt{1 + 6 (6 \alpha_+^2 + 4 \alpha_+ - 1) v_-^2 + 9 v_-^4} \right) \,,
\end{equation}
which relates the fluid velocities on either side of the wall $v_\pm$ as a function of the phase transition strength $\alpha_+$.

Here we comment on the value of the sound speed, which is defined as
\begin{equation}\label{eq:soundspeed}
    c_{s,\pm} \equiv \sqrt{\frac{dp_\pm/dT_\pm}{de_\pm/dT_\pm}} \,.
\end{equation}
Specifically, in the bag model where $a_{\pm}$ are temperature-independent constants, the sound speed is also a constant $c_{s,\pm} = 1/\sqrt{3}$.
While for the general definitions \cref{eq:general_def_a_ep}, $a_{\pm}$ become temperature-dependent, and so does the sound speed. The deviation from $1/\sqrt{3}$, however, is typically not too large. Sound speed plays a crucial role in the fluid profile induced by the bubble expansion, as one will see in Sec.~\ref{sec:type}.
%%%%%%%%%%%%%%%%%%%%%%%%%%%%%%%%%%%%%%%%%%%%%%%%%%%%%%%%%
\subsection{Fluid Profiles}\label{sec:type}

The matching conditions for the asymptotic fluid velocities $v_\pm$ and temperatures $T_\pm$ are provided by \cref{eq:bagEOSmatching1,eq:bagEOSmatching2,eq:matchcond3}, yet the relationship between these quantities and $v_w$ remains undetermined. As we will see in this subsection, this relationship depends on the global fluid profiles induced by the bubble expansion. In the following, we review the three types of fluid profile, and how $v_w$ enters the matching conditions in each case.
We summarize the results in Table~\ref{tab:det-def-hyb} for the readers' convenience.

To solve for the fluid profiles around the bubble, it is more convenient to work in the cosmic frame where the center of the bubble is at rest.
We define a dimensionless quantity $\xi \equiv R/t$, assuming spherical symmetry,
with $R$ being the radial distance from the bubble center, and $t$ being the time since nucleation. The profile of the disturbed plasma is then described by the local fluid velocity $v_{\rm cf}(\xi)$ and temperature $T(\xi)$.

\begin{figure}[t]
    \centering
    \hspace*{-0.5cm}
    \includegraphics[width=0.9\linewidth]{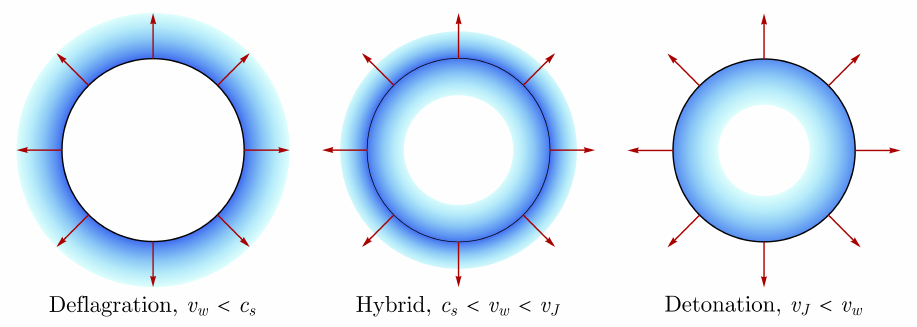}
    \caption{Fluid profiles for deflagration, hybrid, and detonation solutions. The blue shaded region indicates non-zero fluid velocity.}
    \label{fig:det-def-hyb}
\end{figure}

In \cref{app:fluid}, we derive the following master equation for the velocity profile\footnote{\cref{app:fluid} also contains the derivation for the master equation for the temperature profile $T(\xi)$.}
\begin{equation}\label{eq:masterfluid}
    2 \frac{v_{\rm cf}(\xi)}{\xi} = \gamma_{\rm cf}^2 (1 - v_{\rm cf}(\xi) \, \xi) \left( \frac{\mathfrak{v}(\xi, v_{\rm cf}(\xi))^2}{c_s^2(\xi)} - 1 \right) \frac{\partial v_{\rm cf}(\xi)}{\partial \xi} \,,
\end{equation}
where we have defined the Lorentz boost as
\begin{equation}\label{eq:boost}
    \mathfrak{v}(\xi, v) = \frac{\xi - v}{1 - \xi \, v} \,.
\end{equation}
One property of Eq.~\eqref{eq:masterfluid} is that $(\xi = c_s, v_{\rm cf}=0)$ is a stationary point of this equation,
indicating a vanishing fluid velocity at $\xi=c_s$.
Eq.~(\ref{eq:masterfluid}) admits three broad classes of solutions for the fluid profile~\cite{Espinosa:2010hh} --- detonation, deflagration, and hybrid --- which we depict schematically in Fig.~\ref{fig:det-def-hyb}. In the remainder of this subsection, we discuss each of these solutions.

\subsubsection{Detonation}

A detonation solution occurs for fast-moving bubble walls and is characterized by a fluid at rest in front of the wall and a rarefaction wave of the perturbed fluid behind the wall. Since the fluid in front of the bubble wall remains unperturbed, in the wall frame we can identify
\begin{equation}
    v_+ = v_w \,, \,\,\, T_+ = T_n \,,
\end{equation}
where $T_n$ is the nucleation temperature,
defined as the temperature when the bubble nucleation rate reaches the Hubble rate.
$v_-$ and $T_-$ can then be directly solved from~\cref{eq:matchcond1,eq:matchcond2}.

Such a type of fluid profile only exists when the bubble wall moves fast enough so that the disturbed plasma falls behind.
We now derive the allowed lowest value of $v_w$ consistent with a detonation profile.
Working in the cosmic frame, we denote the velocity of a fluid element at a point $\xi$ in the wave profile of the rarefaction wave behind the wall by $v_{\rm cf}(\xi)$, with $\xi < v_w$.
This velocity is largest right behind the wall at $\xi_w=v_w$, where $v_{\rm cf}(\xi_w) = \mathfrak{v}(v_w, v_-)$, and decreases monotonically as one moves further towards the interior of the bubble, i.e. along negative $\xi$ direction, eventually vanishing at $\xi = c_s^-$.
In other words, $v_w \geq c_s^-$ is a necessary condition for a rarefaction wave to exist behind the wall. A consistent solution for $v_{\rm cf}(\xi)$ in the regime $c_s^- < \xi < \xi_w$ requires $\partial_\xi v_{\rm cf} > 0$.
Considering Eq.~(\ref{eq:masterfluid}) with such a requirement, we see that the boosted fluid velocity must exceed the sound speed in the bubble interior, $\mathfrak{v}(\xi, v_{\rm cf}(\xi)) \geq c_s^-$.
At the location of the wall, this implies $v_- \geq c_s^-$.
This condition allows us to define the \textit{Jouguet velocity} $v_J$ as the smallest value of $v_w$ compatible with a detonation profile,
\begin{align}
    \label{eq:vJ def}
    v_J\equiv v_w ~\text{ such that }~ v_- = c_s^- \text{ under detonation profile} \,.
\end{align}
For values of $v_- \geq c_s^-$, we have $v_w \geq v_J$ and the detonation profile is allowed.

\subsubsection{Deflagration}

For $v_w<c_s^-$, we have deflagration solutions, for which the plasma behind the wall is at rest and a shockwave precedes the wall.
The upper bound on $v_w$ comes from the following counter-argument: if $v_w > c_s^-$, a solution with $dv_{\rm cf}(\xi)/d\xi > 0$ would be allowed in the range $c_s^- < \xi < \xi_w$, as can be seen from Eq.~\eqref{eq:masterfluid}.
Therefore, a rarefaction wave would exist, and the bubble wall would no longer be deflagration-type.
Hence, when $v_w < c_s^-$, the plasma behind the wall is at rest, and we can identify
\begin{equation}
    v_- = v_w \,.
    \label{eq:vminus_def}
\end{equation}
Moreover, in the deflagration case,
the thermodynamic quantities evolve in the perturbed region in front of the wall.
This perturbation starts from the bubble wall and ends at a distance in front of it, known as the shock wave front, whose location is denoted as $\xi_{\rm sh}$.
The fluid outside of $\xi_{\rm sh}$ is at rest, and we have
\begin{equation}
    T_{\rm sh,+} = T_n \,,
    \label{eq:Tshock}
\end{equation}
where the $+$ labels the symmetric side of the shock wave front.

The boundary conditions~\cref{eq:vminus_def,eq:Tshock} are not at the same location.
To solve for $T_-$ and $v_+$, $T_+$, we need to connect the bubble wall and the shock wave front.
As before, we move to the cosmic frame and analyze the fluid profile $v_{\rm cf}(\xi)$. The fluid velocity is fastest immediately outside the wall, where $v_{\rm cf}(\xi_w) = \mathfrak{v}(\xi_w, v_+)$.
Moving outward along $\xi$, it decreases monotonically until arriving at the shock wave front, where the temperature and velocity of the plasma are discontinuous.
The decreasing $v(\xi)$ implies that $\mathfrak{v}(\xi, v_{\rm cf}(\xi)) \leq c_s^+$, as can be inferred from Eq.~\eqref{eq:masterfluid}.
At the wall, this becomes $v_+ \leq c_s^+$.
The matching conditions \cref{eq:mc1,eq:mc2} can be directly applied at the shock wave front, with no discontinuity of the scalar field value.
In terms of the cosmic frame velocities, we identify $v_{\rm sh,-}=\mathfrak{v}(\xi_{\rm sh}, v_{\rm cf}(\xi_{\rm sh}))$, and $v_{\rm sh,+}=\mathfrak{v}(\xi_{\rm sh}, 0)=\xi_{\rm sh}$.
Here, $v_{\rm sh, \pm}$ is the fluid velocity on both sides of the shock wave front, measured in the shock wave frame.
Recall the definition of the sound speed, the first matching condition~\eqref{eq:mc1} becomes
\begin{equation}\label{eq:sh-location}
    \mathfrak{v}(\xi_{\rm sh}, v_{\rm cf}(\xi_{\rm sh})) \xi_{\rm sh} = (c_s^+)^2 \,,
\end{equation}
which is used to determine the location of the shock wave front, $\xi_{\rm sh}$.
The velocity profile $v_{\rm cf}(\xi)$ applied here can be solved together with the temperature profile $T(\xi)$ from~\cref{eq:masterfluid,eq:mastertemp}.
The corresponding boundary condition is $v_{\rm cf}(\xi_w) = \mathfrak{v}(\xi_w, v_+)$ and $T(\xi_w) = T_+$.
Specifically, $T(\xi_{\rm sh})$ should be regarded as $T_{\rm sh,-}$, which is related to $T_{\rm sh,+}$ by Eq.~\eqref{eq:mc2}.

Based on the discussion above, the relationship between $v_+$, $T_+$ and $T_{\rm sh,+}$ is established.
Since $T_\pm$ and $v_+$ are still unknown, the following shooting method can be applied to solve for them,
\begin{enumerate}
    \item Set $v_- = v_w$ and make an initial guess for $T_-$.

    \item Solve for $v_+$ and $T_+$ using the matching conditions Eqs.~(\ref{eq:bagEOSmatching1}), (\ref{eq:bagEOSmatching2}).

    \item Solve~\cref{eq:masterfluid,eq:mastertemp} with boundary condition $v(\xi_w) = \mathfrak{v}(\xi_w, v_+)$ and $T(\xi_w) = T_+$ for the velocity and temperature profiles $v(\xi)$ and $T(\xi)$.

    \item Using $v(\xi)$ and $T(\xi)$, solve Eq.~\eqref{eq:sh-location} for the location of the shockwave front. Identify $v_{\rm sh,-} = \mathfrak{v}(\xi_{\rm sh}, v_{\rm cf}(\xi_{\rm sh}) )$ and $T_{\rm sh,-} = T(\xi_{\rm sh})$.

    \item Use the matching conditions at $\xi_{\rm sh}$ to solve for $T_{\rm sh,+}$.

    \item Check whether $T_{\rm sh,+} = T_n$. If not, modify $T_-$ and continue iteration.
\end{enumerate}

\subsubsection{Hybrid}

The remaining range of $v_w$ to be discussed is now $c_s^- < v_w < v_J$, which is known as the hybrid solution, characterized by the existence of both rarefaction wave behind the wall and preceding shock wave between the wall and the shock wave front.
Since deflagration requires $v_- \leq c_s^-$ while detonation requires $v_- \geq c_s^-$, the hybrid solution, as an intermediate range, must require
\begin{equation}
    v_- = c_s^- < v_w\,.
\end{equation}

As in the deflagration case, the nucleation temperature is identified with the temperature at the shock wave front
\begin{equation}
    T_{\rm sh,+} = T_n \,.
\end{equation}
Due to the nature of these boundary conditions, one must again use the shooting method algorithm to determine $T_{\rm sh,+}$. This procedure is completely analogous to the deflagration case except that the initial $v_-$ is set by $c_s^-$. Hybrid solutions are sometimes referred to as ``supersonic deflagrations'' given their similarity to deflagrations whilst having $v_w > c_s^-$.

\begin{table}[h]
    \centering
    \hspace*{-0.2cm}
    \begin{tabular}{|c||c|c|c|}
        \hline
              & Deflagration  & Hybrid              & Detonation  \\
        \hline
        $v_w$ & $v_w < c_s^-$ & $c_s^- < v_w < v_J$ & $v_J < v_w$ \\
        \hline
        \makecell{Boundary                                        \\Conditions} & \makecell{$v_- = v_w$, $T_{\rm sh,+} = T_n$,                                     \\ $\mathfrak{v}(\xi_{\rm sh}, v(\xi_{\rm sh})) \xi_{\rm sh} = (c_s^+)^2$} & \makecell{$v_- = c_s^-$, $T_{\rm sh,+} = T_n$,\\$\mathfrak{v}(\xi_{\rm sh}, v(\xi_{\rm sh})) \xi_{\rm sh} = (c_s^+)^2$} &$v_+ = v_w$, $T_+ = T_n$    \\
        \hline
    \end{tabular}
    \caption{Summary of boundary conditions for deflagration, hybrid, and detonation solutions.}
    \label{tab:det-def-hyb}
\end{table}

\subsection{Determining $v_w$ in LTE}\label{sec:vwinLTE}
\begin{figure}
    \centering
    \includegraphics[width=0.48\linewidth]{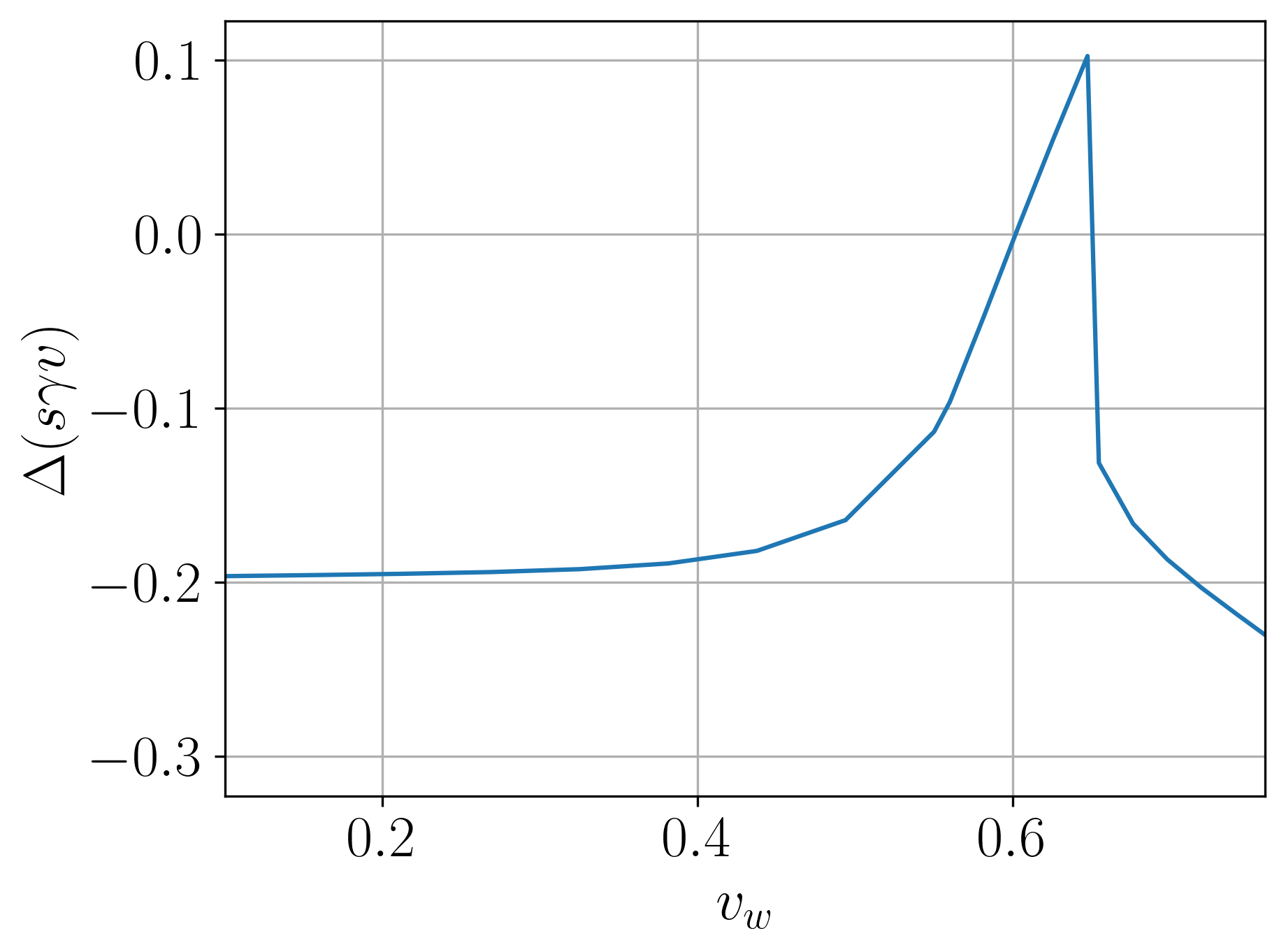}
    \includegraphics[width=0.48\linewidth]{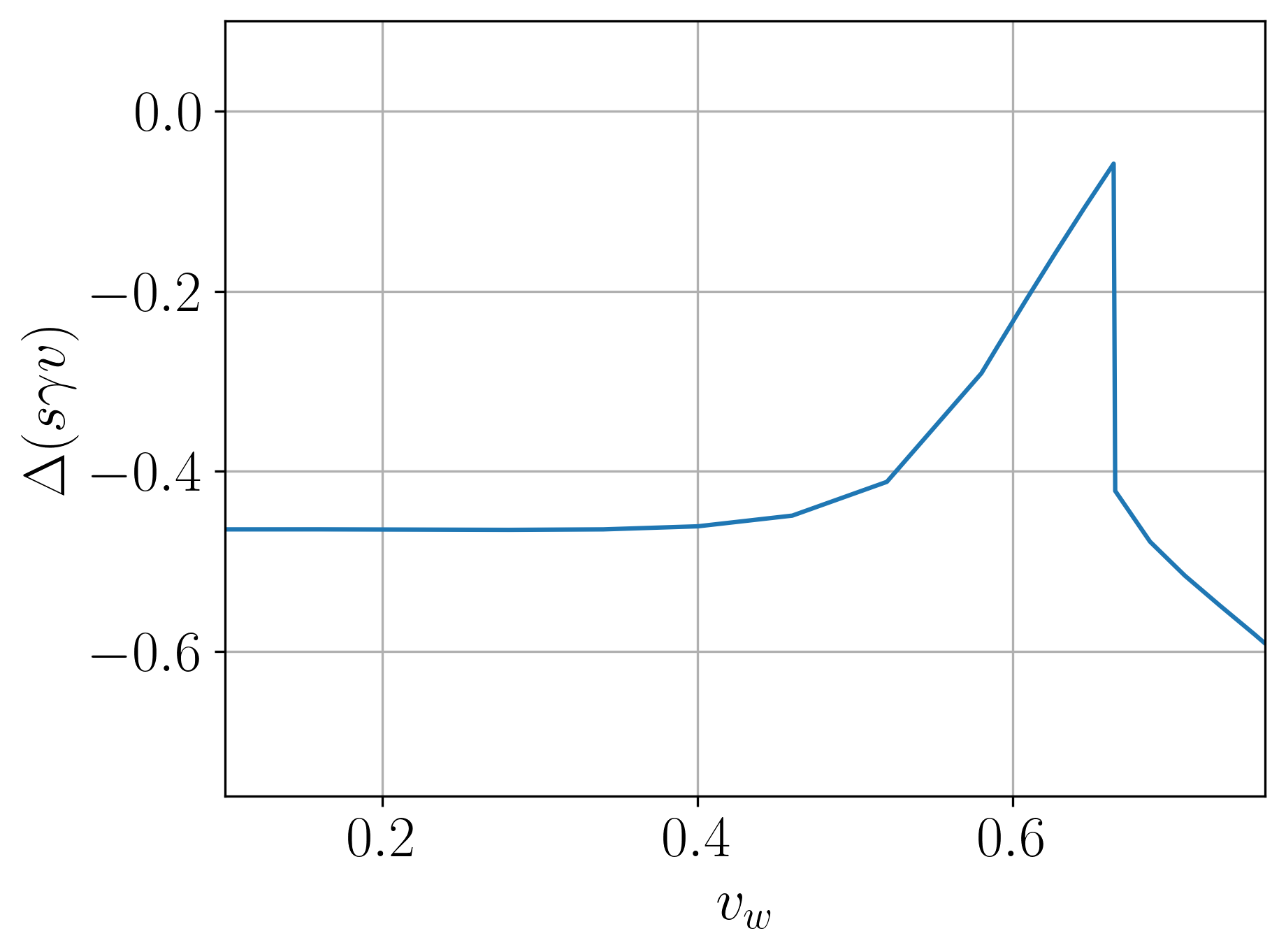}
    \caption{$\Delta (s\gamma v)$ as a function of $v_w$ in arbitrary unit for the $\mathbb{Z}_2$ symmetric real singlet extension model (discussed in \cref{sec:Z2_model}) for two types of bubbles. Left: Deflagration/hybrid-type bubble with a positive peak of $\Delta(s\gamma v)$ at $v_J$. Right: Detonation-type bubble with a negative peak of $\Delta(s\gamma v)$. We truncate the plot at $v_w < 1$, since for $v_w\sim 1$, out-of-equilibrium effects become dominant and LTE is not justified~\cite{Laurent:2022jrs}.}
    \label{fig:ds_vw}
\end{figure}

In summary, for a given $v_w$ and phase transition dynamics, the fluid velocity and temperature near the wall, i.e. $v_\pm$ and $T_\pm$, can be solved based on the two matching conditions,~\cref{eq:matchcond1,eq:matchcond2}, derived from EMT conservation, as well as the boundary conditions depending on the type of solutions as summarized in \cref{tab:det-def-hyb}.
The remaining matching condition demanding entropy conservation in the steady state, $\Delta (s\gamma v) \equiv \left.(s \gamma v)\right|_-^+ = 0$, Eq.~\eqref{eq:matchcond3}, can then be used to determine the true terminal velocity.

Fig.~\ref{fig:ds_vw} shows $\Delta (s \gamma v)$ as a function of $v_w$ in the cases of deflagration/hybrid and detonation for the $\mathbb{Z}_2$ symmetric real singlet extension model, which is used as the test model and discussed in detail in \cref{sec:Z2_model}.
We observe that $\Delta (s \gamma v)$ peaks at $v_w = v_J$ for both deflagration/hybrid and detonation.
This can be intuitively understood recalling that the scalar field EOM is equivalent to entropy conservation in satisfying the condition of EMT conservation, see Eq.~\eqref{eq:entropyproof}.
When $v_w > v_J$, there is no heated plasma in front of the wall, eliminating the friction on the bubble wall and altering the net pressure defined in Appendix~\ref{app:non_LTE}.
This lack of friction, whenever $v_w > v_J$, is responsible for the peak in the total net pressure on the wall at $v_J$.
Such a phenomenon is known as hydrodynamic obstruction~\cite{Konstandin:2010dm}.
If $\Delta (s \gamma v)$ at $v_J$ is positive, there are two solutions of $v_w$ on both sides of $v_J$ satisfying $\Delta (s \gamma v)=0$, see the left panel of Fig.~\ref{fig:ds_vw}.
The solution with smaller $v_w$ of $\Delta (s\gamma v)=0$ gives the true terminal velocity that corresponds to a deflagration/hybrid solution.
If $\Delta (s \gamma v)$ at $v_J$ is negative, there is no solution of $v_w$ satisfying $\Delta (s \gamma v)=0$, as shown in the right panel of Fig.~\ref{fig:ds_vw}.
In this case, no terminal velocity is predicted under LTE, and the bubble wall is detonation-type.
Despite the lack of friction for $v_w>v_J$ in LTE, the out-of-equilibrium effects can still contribute to the friction on the bubble wall, which increases with $v_w$~\cite{Bodeker:2017cim} and only becomes significant when $v_w\to 1$~\cite{Laurent:2022jrs}.
This friction generated from the out-of-equilibrium contribution would eventually balance the driving pressure and stop the bubble wall from accelerating when $v_w\to 1$.
Such an effect is crucial when discussing the scalar field contribution to the GW signal, and will be discussed further in Sec.~\ref{sec:scalar GW}.

We design the following algorithm utilizing the entropy conservation condition first considered in~\cite{Ai:2021kak,Ai:2023see,Ai:2024shx} as:
\begin{enumerate}
    \item Start from a high value of $v_w \to 1$. Solve for $v_-$ and $T_-$ assuming a detonation solution. Decrease $v_w$ until we find $v_- = c_s^-$. Identify such $v_w$ as $v_J$.
    \item Compute the peak of $\Delta (s \gamma v)$ at $v_w \to v_J$ assuming hybrid-type bubble wall. If $\Delta (s \gamma v) < 0$ at this peak, conclude the bubble is detonation-type. If not, the bubble wall is deflagration/hybrid type, and proceed to the next step.
    \item Solve for the true terminal velocity $v_w$ between 0 and $v_J$ satisfying $\Delta (s \gamma v) = 0$ using the deflagration or hybrid boundary conditions as appropriate.
\end{enumerate}

It is worth noting that some recent studies taking the time dependency of the wall velocity into account find that in a large region of parameter space, the steady-state solution of the fluid profile is not fully established during deflagration/hybrid wall expansion~\cite{Krajewski:2024gma,Krajewski:2024zxg}.
Consequently, bubble walls are more likely to enter the detonation regime.
The above recent works offer a challenge to results using the LTE approximation in the existing literature, and we intend to explore this issue in more detail in a future work.

%%%%%%%%%%%%%%%%%%%%%%%%%%%%%%%%%%%%%%%%%%%%%%%%%%%%%%%%%
%%%%%%%%%%%%%%%%%%%%%%%%%%%%%%%%%%%%%%%%%%%%%%%%%%%%%%%%%
\section{Wall Velocity Calculation in Benchmark Models}
\label{sec:vw computation}
In this section, we calculate the terminal wall velocities under LTE for two representative models using the machinery developed in the previous section.

The first model is a $\mathbb{Z}_2$-symmetric real singlet scalar extension of the SM, which is regarded as the most minimal extension capable of rendering a strong first-order EWPT, and has been studied extensively in the literature~\cite{McDonald:1993ey,Choi:1993cv,Espinosa:2007qk,Profumo:2007wc,Espinosa:2011ax,Cline:2013gha,Damgaard:2015con,Kurup:2017dzf,Chiang:2018gsn,Carena:2019une,Friedlander:2020tnq}.
In particular, Ref.~\cite{Laurent:2022jrs} has studied the terminal wall velocity under the general conditions including the out-of-equilibrium effect (see \cref{app:non_LTE} for a review).
Hence, we study this model as a test case to verify explicitly the validity of the LTE approximation and the entropy conservation condition approach, which will then be used for a more complicated scenario.

We next turn to a model with CP violation sourced from the dark sector (henceforth, DarkCPV), studied in Refs.~\cite{Carena:2018cjh,Carena:2019xrr,Carena:2022qpf}.
This model has been shown to have a broad parameter region yielding a strong first-order EWPT.
Using the LTE approximation and entropy conservation approach, we compute the terminal bubble wall velocity in the DarkCPV model.
This will allow us to examine the impact of a first principle calculation of $v_w$ on the results for the baryon asymmetry and the strength of the GW signal.

%%%%%%%%%%%%%%%%%%%%%%%%%%%%%%%%%%%%%%%%%%%%%%%%%%%%%%%%%
\subsection{Real Singlet Scalar Extension with $\mathbb{Z}_2$ Symmetry}
\label{sec:Z2_model}
Consider supplementing the SM with a real scalar singlet $S$ and a discrete $\mathbb{Z}_2$ symmetry under which $S \rightarrow - S$ while the rest of the SM fields remain unchanged. Letting $H$ be the SM Higgs doublet and working in the unitary gauge, the tree-level potential of the scalar sector reads~\cite{McDonald:1993ey,Choi:1993cv,Espinosa:2007qk,Profumo:2007wc,Espinosa:2011ax,Cline:2013gha,Damgaard:2015con,Kurup:2017dzf,Chiang:2018gsn}
\begin{equation}
    V_{\rm tree} = - \mu_H^2 (H^\dagger H)^2 + \lambda_H (H^\dagger H)^4 + \frac{\mu_S^2}{2} S^2 + \frac{\lambda_S}{4} S^4 + \frac{\lambda_{SH}}{2} S^2 (H^\dagger H) \,.
    \label{eq:Vtree_Z2}
\end{equation}
The finite-temperature effective potential $V_{\rm eff}$ is the sum of this tree level potential with the Coleman-Weinberg and finite-temperature corrections, $V_{\rm eff} = V_{\rm tree}+V_{\rm CW} + V_T$, see Appendix~\ref{app:singletscalar} for details.
Among the five parameters in the tree-level potential, two of them are fixed by the SM Higgs mass and VEV, while the other three remain free.
In the following, we parameterize the model as a function of: $\lambda_{S}$, $\lambda_{SH}$, and the physical mass of the real scalar singlet $m_S$.

\begin{figure}[t]
    \centering
    \includegraphics[width=\linewidth]{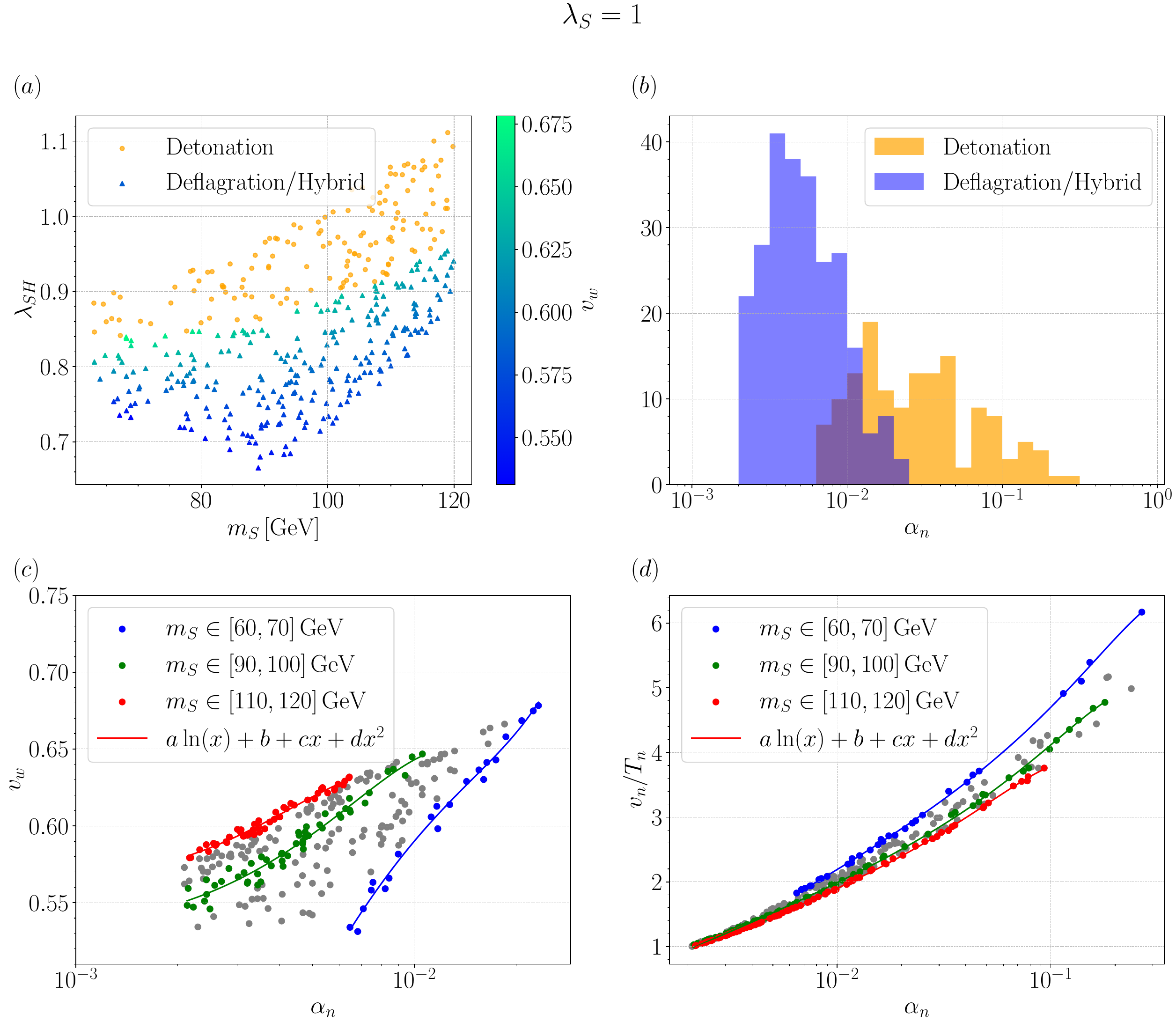}
    \caption{Strong first-order EWPT points and wall velocities calculated under the LTE approximation, using the entropy conservation approach for the $\mathbb{Z}_2$ symmetric real scalar singlet extended SM. We fix $\lambda_S = 1$, and vary $\lambda_{SH}$ and $m_S$ in the scanning. Panel (a) shows detonation and deflagration/hybrid solutions in the $m_S$-$\lambda_{SH}$ plane, with wall velocities depicted by the color gradient. Panel (b) shows the histogram of $\alpha_n$ for the detonation and deflagration/hybrid solutions. Panel (c) shows all the deflagration/hybrid points in the $\alpha_n$-$v_w$ plane. The points highlighted in blue, green, and red have one of the model parameters $m_S$ restricted to a narrow range as indicated by the legend. The solid lines of the same colors are the fitting of the scatter points with the function $f(x)=a\ln(x)+b+cx+dx^2$, with fitted parameters listed in \cref{tab:fit_z2}. Panel (d) shows all the strong first-order EWPT points (in gray), including detonation bubbles, in the $\alpha_n$-$v_n/T_n$ plane. The blue, green, and red points and lines have the same meaning as those in panel (c).}
    \label{fig:scatter_z2}
\end{figure}

To verify explicitly the validity of the LTE approximation and the entropy conservation approach, we use this formalism to compute $v_w$ and compare the results with those in Ref.~\cite{Laurent:2022jrs}.
\cref{fig:scatter_z2} presents the results for 394 benchmark points that have strong first-order EWPT, with fixed $\lambda_S=1$ and varying $m_S$ and $\lambda_{SH}$.
In the parameter region being considered, the EWPT proceeds in two steps in the $(h, S)$ field configuration space: $(0,0)\to(0,w')\to(v_n,0)$. \fig{scatter_z2}{a} shows the detonation and deflagration/hybrid solutions in the $m_S$-$\lambda_{SH}$ plane, with the wall velocities of the deflagration/hybrid points depicted by the color gradient. The wall velocity falls into the range from $\sim 0.53-0.68$, slightly narrower than that in Ref.~\cite{Laurent:2022jrs} due to lower statistics\footnote{Our results, similar to Ref.~\cite{Laurent:2022jrs}, are at variance with simple approximations yielding a very small value of $v_w$, e.g. $v_w \simeq (T_c - T_n)/T_n$ \cite{Asadi:2021pwo}.}.
~\fig{scatter_z2}{b} shows the histogram of $\alpha_n$ for the detonation and deflagration/hybrid solutions.
We adopt the $\alpha_n$ widely used in the literature as given below, which matches \cref{eq:generalalpha} at $T_n$, ignoring the temperature difference between the two phases:
\begin{align}
    \label{eq:alpha_n}
    \alpha_n=  \left.\frac{1}{g_\ast T^4 \pi^2/30} \left(\Delta V_{\rm eff} - \frac{T}{4} \frac{\partial \Delta V_{\rm eff}}{\partial T}\right) \right|_{T_n}
\end{align}
where $g_\ast=106.75$ is the number of the light degrees of freedom in the SM and $\Delta V_{\rm eff}$ refers to the difference of $V_{\rm eff}$ between the false vev and the true vev.
Comparing \fig{scatter_z2}{a} and \fig{scatter_z2}{b} with Fig.~3 and 4 of Ref.~\cite{Laurent:2022jrs} (see also Fig.~5 of Ref.~\cite{Cline:2021iff}), we see that the distribution of the detonation and deflagration/hybrid solutions displays the same pattern.
This adds credence to the claim that the leading-order effects are already present in LTE. Most importantly, our results prove that the entropy conservation algorithm provides a reliable method of computing $v_w$.
In \fig{scatter_z2}{c} and \fig{scatter_z2}{d}, we show scatter plots in the $\alpha_n$-$v_w$ plane (for the deflagration/hybrid points), and in the $\alpha_n$-$v_n/T_n$ plane (for all the benchmark points, same as \fig{scatter_z2}{a} and \fig{scatter_z2}{b}).
We observe positive correlations between $\alpha_n$ and $v_w$, $\alpha_n$ and $v_n/T_n$, respectively, although with two varying parameters ($m_S$, $\lambda_{SH}$), there is no 1-to-1 correspondence between them in either case.
When one parameter is fixed, or restricted to a narrow range, $\alpha_n$ (nearly) uniquely determines $v_w$ and $v_n/T_n$. In \fig{scatter_z2}{c} and \fig{scatter_z2}{d}, the blue, green, and red points have $m_S\in[60,70]\,\gev$, $m_S\in[90,100]\,\gev$ and $m_S\in[110,120]\,\gev$, respectively, with $\lambda_{SH}$ varying in the range illustrated in \fig{scatter_z2}{a}. The blue, green, and red lines are the fitting of the corresponding points with the function $f(x)=a\ln(x)+b+cx+dx^2$, with the fitted parameters summarized in \cref{tab:fit_z2}.
Although the fitting coefficient accompanying the $\ln(x)$ term is rather small, we emphasize that this term dominates in the low-$\alpha_n$ regime and therefore is essential for the fit.

\begin{table}[h]
    \centering
    \begin{tabular}{|c|c|c|}
        \hline
        $m_S$ range $(\rm GeV)$ & $(a,b,c,d)$ for $v_w=f(\alpha_n)$ & $(a,b,c,d)$ for $v_n/T_n=f(\alpha_n)$ \\
        \hline
        $60-70$                 & $(0.27, 2.00, -22.83, 382.07)$    & $(0.78, 5.68, 9.59,-14.60)$           \\
        \hline
        $90-100$                & $(-0.016, 0.40, 27.24, -1021.60)$ & $(0.59,   4.60,  11.48, -26.61)$      \\
        \hline
        $110-120$               & $(-0.028, 0.34, 36.93, -2114.66)$ & $(0.52,   4.13,  16.41, -78.95 )$     \\
        \hline
    \end{tabular}
    \caption{Fitted parameters for $v_w$ and $v_n/T_n$ as function $f(\alpha_n)=a\ln(\alpha_n)+b+c\,\alpha_n+d\,\alpha_n^2$ for $m_S$ in the given ranges.}
    \label{tab:fit_z2}
\end{table}

We would like to further comment on the fact that \fig{scatter_z2}{b} shows a notable overlap between the detonation and deflagration/hybrid solutions. This is at variance with the results of the analytic treatment of Ref.~\cite{Ai:2024shx}, which demonstrate the existence of a critical value of $\alpha_n^{\rm max}$ defining the boundary between the deflagration/hybrid and detonation regimes.
The derivation assumes the bag equation of state, and relies on the fact that the height of the pressure peak is a monotonic function of $\alpha_n$.
It solves for the critical value $\alpha_n^{\rm max}$ as the value at which the height of the peak in \cref{fig:ds_vw} is at zero.
The result is an expression for $\alpha_n^{\rm max}$ which depends solely on the number of light degrees of freedom in the model.
Applying this formula to the $\mathbb{Z}_2$-symmetric scalar singlet extension, one finds the critical value $\alpha_n^{\rm max} \simeq 0.05$.
However, \fig{scatter_z2}{b} shows a more complicated pattern, implying a deviation from the bag equation of state.
Nevertheless, the results in \cite{Ai:2024shx} still provide guidance for the maximal value of $\alpha_n$ for deflagration/hybrid solutions in EWPT.

%%%%%%%%%%%%%%%%%%%%%%%%%%%%%%%%%%%%%%%%%%%%%%%%%%%%%%%%%
\subsection{Dark CP-Violating Model}

We now turn to the DarkCPV model of Refs.~\cite{Carena:2018cjh,Carena:2019xrr,Carena:2022qpf}, which achieves EWBG whilst evading the strong constraints on electric dipole moments by confining the CP violation to the dark sector.
In this model, the scalar sector is extended by a complex singlet scalar $\SC = s + ia$, which couples to the SM Higgs to facilitate a strong first-order EWPT.
CP violation is sourced by the complex Yukawa coupling between $\SC$ and the chiral dark fermions $\chi_L$ and $\chi_R$\footnote{$\chi_{L,R}$ can also serve as a good dark matter candidate.},
\begin{align}
    \mathcal{L} \supset \bar{\chi}_L (m_0 + \lambda_c \SC) \chi_R + \rm h.c,
    \label{eq:dark_yukawa}
\end{align}
where the Yukawa coupling $\lambda_c$ is a complex number.
To ensure the complex phase cannot be rotated away, the tree-level scalar potential is extended by a complex coupling term $\kappa_\SC^2 \SC^2$ and is given by,
\begin{align}
    V_{\rm tree} = & \lambda_H (|H|^2 - v_H^2)^2 + \lambda_\SC (|\SC|^2 - v_{\SC}^2)^2 + \lambda_{\mathbb{S}H} |\mathbb{S}|^2 |H|^2
    + \kappa_\mathbb{S}^2 \mathbb{S}^2 + \rm h.c.
\end{align}
In our parameterization, we use the freedom of field redefinition to fix $m_0$ and $\kappa_\mathbb{S}^2$ to be real, leaving $\theta \equiv \arg(\lambda_c)$ as the only remaining CP-violating phase.
Imposing the SM Higgs mass and VEV ($\lambda_H=0.129$, $v_H=246\,\gev$), the model can be parameterized by 5 free parameters in the scalar sector and 3 parameters for the dark fermion:
\begin{align}
     & \text{Scalar sector: }\lambda_\mathbb{S},\,\lambda_{\mathbb{S}H},\,v_\SC' \equiv \sqrt{v_\SC^2 + 2 \kappa_\SC^2/\lambda_\SC},\,m_s,\,m_a, \\
     & \text{Dark fermion: }m_0,\,|\lambda_c|,\,\theta,
\end{align}
where $m_s$ and $m_a$ are the physical masses of the real and imaginary parts of $\SC$.
Appendix~\ref{app:singletscalar} shows the details of the effective potential, which leads to a one-step phase transition in the $(h, s, a)$ field configuration space: $(0,s_\mathbb{S},a_{\mathbb{S}})\to (h_{\rm EW}, s_{\rm EW}, a_{\rm EW})$.
The non-restoration of the $\mathbb{Z}_2$ symmetry for $\mathbb{S}$ at high temperature is due to the thermal corrections from the dark Yukawa coupling given by \cref{eq:dark_yukawa} (see discussions in \cite{Carena:2022qpf}).

\begin{figure}
    \centering
    \includegraphics[width=\linewidth]{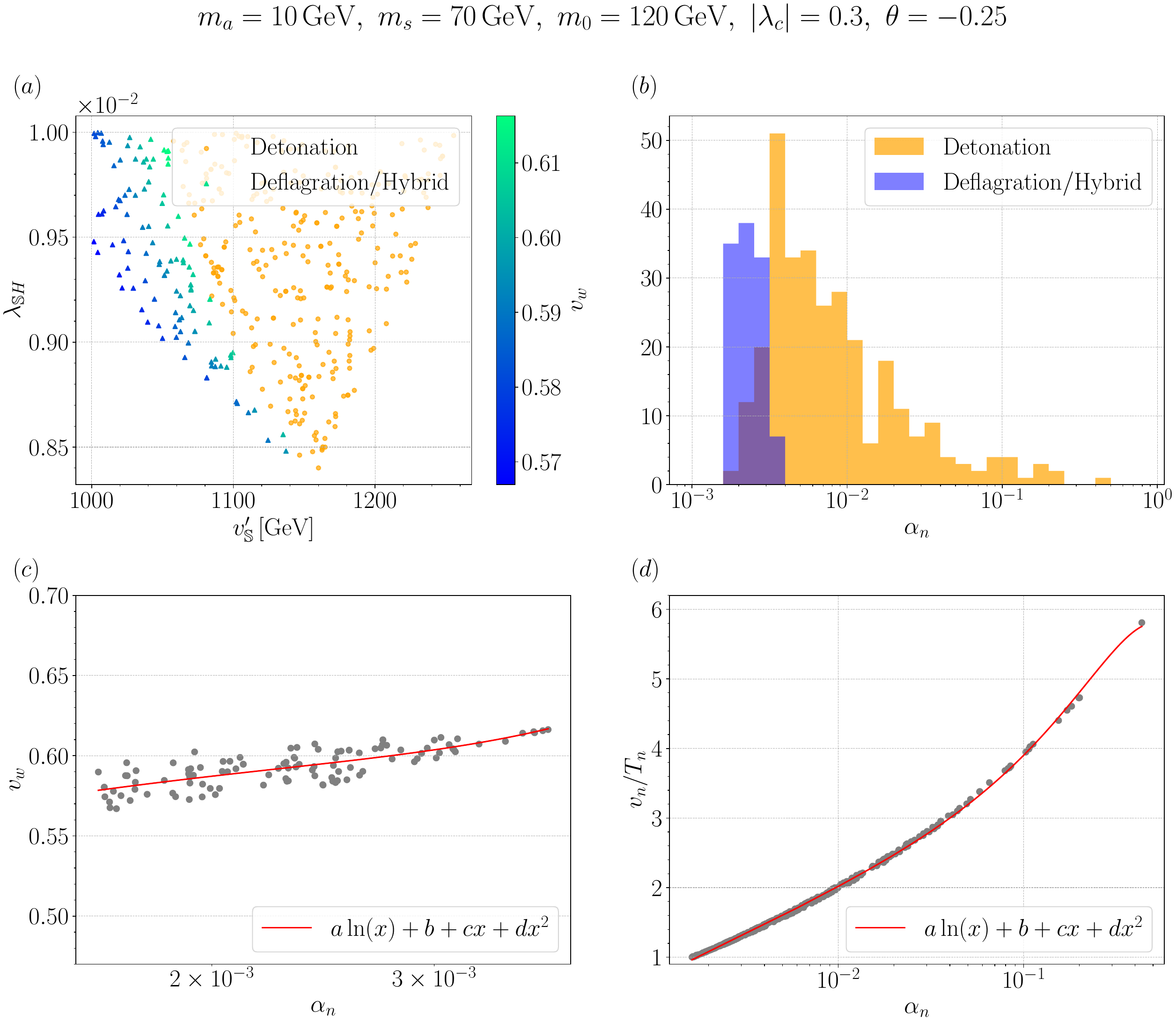}
    \caption{Strong first-order EWPT points and $v_w$ calculated under LTE using the entropy conservation approach for the DarkCPV model. There are two free parameters varying in the scanning: $v'_\mathbb{S}$ and $\lambda_{\SC H}$. Other parameters are fixed as given in the plot title. (a) Detonation and deflagration/hybrid points in the $v'_\SC$-$\lambda_{\SC H}$ plane. Wall velocities for the deflagration/hybrid points are depicted by the color gradient. (b) Histogram of $\alpha_n$ for the detonation and deflagration/hybrid solutions. (c) $\alpha_n$ and $v_w$ of the deflagration/hybrid points. The red line is a fitting of the scatter points with the function indicated by the legend. (d) $\alpha_n$ and $v_n/T_n$ of all the strong first-order EWPT points. The red line is again a fitting with the same function as in panel (c).}
    \label{fig:dcpv_scatter}
\end{figure}

We now perform a numerical scanning for the parameters giving a strong first-order EWPT, and compute $v_w$ for these points in the LTE approximation and using the entropy conservation approach.
Based on the previous subsection, we expect this method to yield a reliable result.
As shown in Ref.~\cite{Carena:2022qpf}, collider bounds from Higgs exotic decays require $m_s > m_h/2$ in order to allow for a SFOPT.
In addition, considering that the observed dark matter relic density is mainly determined by the annihilation channel $\bar{\chi} \chi$ into scalars,
$m_0$ should be appreciably larger than $m_a$ or $m_s$.
On the other hand, a too heavy $m_0$ leads to too strong baryon number washout based on the updated baryon number generation calculation discussed in \cref{sec:BAU}.
Considering all the above,
we choose the parameter region for which $m_a<m_h/2$, $m_s>m_h/2$, $m_0 > m_{a,s}$, and $\lambda_{SH} \lesssim 0.01$, the latter to avoid collider bounds from Higgs invisible decay.

\cref{fig:dcpv_scatter} shows 417 benchmark points with a strong first-order phase transition for fixed values of $m_a=10\,\gev$, $m_s=70\,\gev$, $m_0=120\,\gev$, $|\lambda_c|=0.3$, $\theta=-0.25$.
Counterterms are added to cancel the Coleman-Weinberg correction so that the tree-level relation $m_a^2=\lambda_{\SC H}v_H^2/2-\lambda_\SC v_{\SC}'^2$ is fixed.
Only two free parameters then remain in the scanning: $v_{\SC}'$ and $\lambda_{\SC H}$, as depicted in \fig{dcpv_scatter}{a}.
The wall velocities for the deflagration/hybrid walls are in the range $\sim 0.56-0.62$.
Similar to the $\mathbb{Z}_2$ singlet scalar model, a histogram of $\alpha_n$ for the detonation and deflagration/hybrid solutions is shown in \fig{dcpv_scatter}{b} for the DarkCPV model,
from which we observe a much lower upper bound of $\alpha_n$ for the deflagration/hybrid bubbles at around $4\times 10^{-3}$.
\fig{dcpv_scatter}{c} and \fig{dcpv_scatter}{d} show the deflagration/hybrid points in $\alpha_n$-$v_w$ plane and all the SFOPT points in $\alpha_n$-$v_n/T_n$ plane, respectively.
Due to the collider bounds discussed before, the values of $\lambda_{SH}$ fall into a narrow range.
This resembles the scenario in the $\mathbb{Z}_2$-symmetric real singlet extension model with $m_S$ constrained in a narrow range (the blue, green, and red points in \fig{scatter_z2}{c} and \fig{scatter_z2}{d}).
As before, a fitting with the same fit function $f(x)=a\ln(x)+b+cx+dx^2$ is performed for $v_w$-$\alpha_n$ for all the deflagration/hybrid points, and $v_n/T_n$-$\alpha_n$ for all the strong first-order EWPT points, as shown in \fig{dcpv_scatter}{c} and \fig{dcpv_scatter}{d}.
The fitted parameters are:
\begin{align}
     & v_w=f(\alpha_n):~a=0.15,~b=1.64, ~c=-95.07,~  d=1.05\times 10^{4}.\nonumber \\
     & \frac{v_n}{T_n}=f(\alpha_n):~a=0.54,~  b=4.44,~ c= 8.01,~ d=-9.10.
\end{align}
An interesting observation is that the fitting coefficients for the $\mathbb{Z}_2$ singlet extension model and DarkCPV model are of comparable orders, indicating universal correlation patterns of $\alpha_n$-$v_w$ and $\alpha_n$-$v_n/T_n$ across the singlet scalar extensions of the SM.
However, the significant model dependence of the allowed values of $\alpha_n$ for which deflagration/hybrid bubbles occur, as shown in \fig{scatter_z2}{b} and \fig{dcpv_scatter}{b}, respectively, highlights the importance of a careful treatment of the fluid profiles and wall velocity for each model.

%%%%%%%%%%%%%%%%%%%%%%%%%%%%%%%%%%%%%%%%%%%%%%%%%%%%%%%%%
%%%%%%%%%%%%%%%%%%%%%%%%%%%%%%%%%%%%%%%%%%%%%%%%%%%%%%%%%
\section{Implications for Baryogenesis}\label{sec:BAU}
The wall velocities obtained in the previous section for the real and complex singlet scalar extensions of the SM are in the range $\sim 0.5-0.7$, which are much higher than the commonly assumed value $v_w\sim 0.1$ for EWBG calculations, see, e.g.~\cite{Grzadkowski:2018nbc,Carena:2019xrr,Carena:2022qpf}.
Accordingly, the non-local baryon number generation based on the diffusion mechanism has to be re-examined for wall velocities in this range.
In this section, we compare different computational frameworks for the baryon number generation and examine their validity depending on the range of $v_w$.
We then present the numerical results in the DarkCPV model, demonstrating distinct behaviors for the baryon asymmetry in the proximity of the real value of $v_w$.
%%%%%%%%%%%%%%%%%%%%%%%%%%%%%%%%%%%%%%%%%%%%%%%%%%%%%%%%%
\subsection{$v_w$-Dependence of Baryon Number Generation}
\label{sec:nb_calculation}
In non-local EWBG, which relies on CP-violating particle reflecting off the bubble wall, the behavior of the BAU under an increasing wall velocity has been long believed to involve two competing effects.
On the one hand, a faster wall velocity allows reflected particles to diffuse back into the bubble wall more quickly, reducing the window in which the sphaleron process can convert the CP asymmetry into baryon asymmetry, and thus suppressing the BAU.
On the other hand, higher wall velocity enhances the scattering rate between the plasma particles and the bubble wall, producing a larger CP asymmetry and thereby increasing the BAU.
A detailed understanding of the dependency of BAU on $v_w$ requires solving the transport equation accurately in the wall frame.

Two commonly used frameworks have been developed to formulate the transport equation: the semi-classical approach under the WKB approximation~\cite{Joyce:1994bi,Joyce:1994zn,Joyce:1994zt,Cline:1995dg,Cline:2000nw,Fromme:2006wx}, and the VEV-insertion approximation (VIA) method~\cite{Huet:1994jb,Huet:1995sh,Riotto:1995hh, Riotto:1998zb, Lee:2004we}.
The essential difference between these two methods lies in their calculations of the CP-violating source term that enters the transport equation governing the distribution of particle number densities near the bubble wall.
The VIA method treats the CP-violating mass term induced by the bubble wall as a perturbation that flips chirality, and calculates the source term to leading order in the mass insertion.
Although originally calculations based on the VIA approximation seemed to accommodate EWBG for a large number of models, more recent studies have shown pathologies that call into question its validity~\cite{Kainulainen:2021oqs,Postma:2022dbr}.
Further studies are needed to understand if future improvements in the VIA approximation could lead to reliable results.
% However, this approach has now been disfavored by recent studies in e.g., \cite{Kainulainen:2021oqs,Postma:2022dbr}, which point out inconsistencies xxx.
The semi-classical method, which will be used throughout this work, describes the interaction of the bubble wall as a semi-classical force in the WKB approximation. This approach is only valid in the thick wall regime, where the particle's de Broglie wavelength ($\sim T^{-1}$) is much smaller than the wall width $L_w$\footnote{The real $L_w$ can be computed from first principles, as shown in Appendix~\ref{app:non_LTE}, and has been done for the $\mathbb{Z}_2$ model in Ref.~\cite{Cline:2021iff}. Given the similarity in the range for  $v_w$ for the $\mathbb{Z}_2$  and the DarkCPV models, we shall assume that the wall thickness is in a similar range as in the $\mathbb{Z}_2$ case ($L_w\sim 5/T$). We leave a first-principle calculation of $L_w$ for the DarkCPV model for future work.}.

Earlier calculations for the baryon number generation using the semi-classical method by Cline, Joyce, and Kainulainen \cite{Cline:2000nw}  and  by  Fromme and Huber \cite{Fromme:2006wx} assumed $v_w\ll 1$ and truncated the transport equation to the leading order of $v_w$. We will refer to this framework as the leading order (L.O.) approximation.
More recently, computations by  Cline and Kainulainen~\cite{Cline:2020jre} that keep the full $v_w$ dependence uncover a significantly different behavior of the baryon number as a function of $v_w$. We will refer to this all orders calculation as the full treatment.
In this section, we compute the baryon asymmetry generated in the DarkCPV model using the full treatment of $v_w$. We find striking differences between these results and those obtained using the L.O. $v_w$ approximation, which is only valid for small $v_w$\footnote{A comparison of the L.O. and full treatments has also been performed in Ref.~\cite{Cline:2020jre}.}.
Detailed derivations are provided in Appendix~\ref{sec:app_bau}.

The starting point is the Boltzmann equation in the bubble wall frame, focusing on the region near the wall. Because the curvature of the bubble wall can be ignored in this local area, the Boltzmann equation reduces to a one-dimensional form along the direction perpendicular to the wall ($z$):
\begin{align}
    \label{eq:BAU Boltzmann}
    \left(v_g\partial_z+F_{z}\partial_{k_z}\right)f_i = C_i[f_i, f_j,...],
\end{align}
where we have introduced the group velocity and force in the $z$ direction as, $v_g\equiv \Dot{z}$ and $F_{z}\equiv\Dot{k_z}$, respectively.
$C_i[f_i, f_j,...]$ is the collision term depending on various particle species.
For a fermion with CP-violating complex mass term $m(z)=|m(z)|e^{i\theta(z)\gamma^5}$, taking the WKB approximation, these quantities are given by:
\begin{align}
    v_g & =(\partial_{k_c}E_w)_z \equiv\frac{k_z}{E_w},\label{eq:group_velocity}                                             \\
    F_z & =-\frac{(|m|^2)^\prime}{2E_w} + ss_{k_{0}}\frac{\left(|m|^2\theta^\prime\right)^\prime}{2E_w E_{wz}},\label{eq:Fz}
\end{align}
where $E_w$ is the conserved energy in the wall frame, and $E_{wz}^2\equiv E_w^2-(k_x^2+k_y^2)$.
Throughout this work, we use $^\prime$ to denote $\partial_z$ for all quantities other than the particle distribution function $f$.
The derivative on $f$ will be defined later.
The first equality of \cref{eq:group_velocity} follows from the classical Hamiltonian equation for the WKB wave packet, with $k_c$ being the canonical momentum. The second equality of \cref{eq:group_velocity} can be taken as the definition of the kinetic momentum in the $z$ direction--$k_z$.
The CP violation effect is induced by the second term of $F_z$, with $ss_{k_{0}}=-1$ for the left-handed fermion with $k_z>0$, and $ss_{k_{0}}=+1$ for its CP conjugate--the right-handed anti-fermion.
In presence of a spatially varying phase profile $\theta(z)$, $E_w$ is given by the physical momentum as,
\begin{align}
     & E_w\approx E_0 - ss_{k_0}\frac{|m|^2\theta^\prime}{2E_0 E_z}\equiv E_0+ss_{k_0}\Delta E,\label{eq:Ew_and_E0} \\
     & {\rm with}~~E_0\equiv\sqrt{\mathbf{k}^2+m^2},~~E_z\equiv\sqrt{k_z^2+m^2}.
\end{align}
This is analogous to the conserved energy of a charged particle in an electromagnetic field.
Substitute these into \cref{eq:group_velocity} and \cref{eq:Fz} and keep the leading-order terms in spatial derivative, we get
\begin{align}
     & v_g\approx\frac{k_z}{E_0}\left(1 + s s_{k_0}\frac{|m|^2\theta'}{2 E_0^2 E_z}\right),\label{eq:vg_E0}                                                   \\
     & F_z\approx -\frac{(|m|^2)'}{2 E_0}+s s_{k_0}\left[\frac{(|m|^2\theta')'}{2 E_0 E_z} - \frac{|m|^2(|m|^2)'\theta'}{4 E_0^3 E_z}\right].\label{eq:Fz_E0}
\end{align}

It is difficult to solve the Boltzmann equation directly in the presence of the collision term, which is a non-local integral over the phase space. To proceed, one usually needs to take an ansatz.
The methods to be reviewed and compared here, i.e. the leading order (L.O.) approximation and all orders full treatment, are based on the ansatz of a perturbed equilibrium distribution boosted to the wall frame,
\begin{align}
    \label{eq:ansatz general}
    f = \frac{1}{e^{\beta\left[\gamma_w(E_w+v_w k_z)-\mu(z)\right]}\pm 1} +\delta f(k,z).
\end{align}
Here $\gamma_w=1/\sqrt{1-v_w^2}$ is the Lorentz boost factor.
In this ansatz, the departure from chemical equilibrium is encoded in $\mu(z)$, and the departure from kinetic equilibrium is encoded in $\delta f(k,z)$, which satisfies the following condition so that it does not affect the local particle density
\begin{equation}
    \int\!d^3 k\,\delta f=0.
\end{equation}
The problem then reduces to solving the differential equation for $\mu$ and $\delta f$, which is non-linear and hard to solve upon substituting \cref{eq:ansatz general} into the Boltzmann equation.
To proceed, one could expand over $\mu$ and $\delta f$, and keep the linear terms,
as detailed in \cref{sec:app_bau}.
In the following, we compare two linearization schemes that keep different orders of $v_w$ in the coefficients\footnote{Note that there is another commonly employed ansatz modeling the out-of-equilibrium distribution in a perfect fluid form \cite{Moore:1995si,Dorsch:2021nje}, predicting a singularity at the sound speed; whether this singularity is physical or not is under debate.
    In this work, however, we focus on the ansatz in Eq.~\eqref{eq:ansatz general}.}.

One way to solve the linearized differential equation is to take moments over momentum space, weighting by $1$ and $k_z/E_0$ respectively. This gives two independent moment equations for two variables, the chemical potential $\mu$, and a new variable $u\equiv\langle\frac{k_z}{E_0}\delta f\rangle$, with the angle bracket defined as
\begin{align}
    \langle X\rangle\equiv\frac{\int\! d^3 k\, X}{\int\! d^3 k\, f_0^\prime} \,.
    \label{eq:norm_moment}
\end{align}
Here $f_0$ is the equilibrium distribution boosted to the wall frame. In the full treatment, the relativistic Lorentz boosting is retained, and the distribution function $f_0$ is given by
\begin{align}
    \label{eq:equilibrium CK}
    f_{0w}=\frac{1}{e^{\beta\gamma_w(E_0+v_w k_z)}\pm 1} \,.
\end{align}
Correspondingly, $f_{0w}'=\partial f_{0w}/\partial({\gamma_{w}E_0})$. The moment equations can be written in the matrix form as
\begin{equation}
    \label{eq:transport}
    A\omega^\prime + B\omega=S + C \,,
\end{equation}
with $\omega=(\mu,u)^T$, which in general can be decomposed into the CP-even and CP-odd contributions as $(\mu,u)=(\mu_e,u_e)+s_{k_0}(\mu_o,u_o)$. The transport equations for the CP-even and CP-odd components decouple after linearization as discussed in \cref{sec:app_bau}.  In this section, we focus on the transport equation for the CP-odd terms, which is relevant for baryogenesis.
In the all orders treatment that keeps the full dependence on $v_w$, the matrices of the derivative coefficients $A$, the linear coefficients $B$, the source terms $S$, and the collision terms $C$ take the following forms
\begin{align}
    \label{eq:coefficient CK20}
     & A= \begin{pmatrix}
              -D_1 & 1     \\
              -D_2 & - v_w
          \end{pmatrix},\quad B=(|m|^2)^\prime\begin{pmatrix}
                                                  v_w\gamma_w Q_1 & 0       \\
                                                  v_w\gamma_w Q_2 & \bar{R}
                                              \end{pmatrix},\notag                                                                                                         \\
     & S=\begin{pmatrix}
             S_1 \\ S_2
         \end{pmatrix}, \quad{\rm with}~~S_\ell\equiv v_w\gamma_w\left[(|m|^2\theta^\prime)^\prime Q_\ell^{8o} - (|m|^2)^\prime |m|^2\theta^\prime Q_\ell^{9o}\right],\nonumber \\
     & C=\begin{pmatrix}
             C_1 \\ -\Gamma_i^{\rm tot}u-v_w C_1
         \end{pmatrix},\quad{\rm with}~~C_1=K_0\sum_c \Gamma_{i}^c\sum_{j}s_{j}^c\frac{\mu_j^c}{T} \,.
\end{align}
Here, $\Gamma_{i}^c$ and $\Gamma_i^{\rm tot}$ are the inelastic interaction rates for species $i$ in channel $c$ and in total, respectively, where $s_{j}^c=+1(-1)$ if the species $j$ is in the initial (final) state in channel $c$. The coefficients appearing in the moment equations above are given by,
\begin{align}
     & D_\ell\equiv\langle\left(\frac{k_z}{E_0}\right)^\ell f_{0w}'\rangle,\quad Q_\ell\equiv\langle\left(\frac{k_z^{\ell-1}}{2E_0^\ell}\right)f_{0w}''\rangle,\quad \bar{R}=\frac{\int d^3 k\,\left(\frac{1}{2k_z E_0}\right)f_{0w}}{\int d^3 k\, f_{0w}},\notag                              \\
     & Q_\ell^{8o}\equiv\langle\frac{s_k k_z^{\ell-1}}{2E_0^{\ell}E_z}f_{0w}'\rangle,\quad  Q_\ell^{9o}\equiv\langle\frac{s_k k_z^{\ell-1}}{2E_0^{\ell+1}E_z}\left(\frac{1}{E_0}f_{0w}'-\gamma_w f_{0w}''\right)\rangle,\quad K_0\equiv-\frac{\int d^3 k f_{0w}}{\int d^3 k f_{0w}^\prime} \,,
    \label{eq:ck20_coeffs_2}
\end{align}
where $s_k={\rm sign}(k_z)$ denotes the sign of the $z$-direction momentum. The source terms for the CP-odd transport equation are defined in \cref{eq:S_cpodd} and describe the CP-violating effects for the left-handed particles, which are non-vanishing only for non-trivial $\theta'$.

\cref{eq:transport,eq:coefficient CK20} are the final equations to solve in the all orders treatment. Note that these equations have no singularity in $v_w$ and behave smoothly for all $0 < v_w < 1$.
In contrast, earlier methods assuming small wall velocity which took only the leading order term in an expansion in $v_w$ frequently encounter singular behavior at $v_w < 1$. We discuss the origin of this singularity in more detail in \cref{sec:app_bau}. To summarize, in the L.O. approximation, the Lorentz transformation is approximated as the Galilean transformation at small $v_w$, such that $f_0^{\rm L.O.}=(e^{\beta E_0}\pm 1)^{-1}$. The derivative coefficient matrix $A$, the source term $S$, and the collision term $C$ are given by,
\begin{align}
    \label{eq:coefficient CJK}
    A_{\rm L.O.} & =\begin{pmatrix}
                        -v_w                                         & 1   \\
                        -\langle\frac{k_z^2}{E_0^2}f_0^\prime\rangle & v_w
                    \end{pmatrix},\quad
    B_{\rm L.O.}=\mathbf{0}, \nonumber                                                                                      \\
    S_{\rm L.O.} & =\begin{pmatrix}
                        0 \\ -v_w\langle\frac{k_z}{E_0}F_{z}f_0^\prime\rangle
                    \end{pmatrix},\quad C_{\rm L.O.}= \begin{pmatrix}K_0\sum_c \Gamma_{i}^c\sum_{j}s_{j}^c\frac{\mu_j^c}{T} \\
                    -\Gamma_i^{\rm tot}u
                    \end{pmatrix}.
\end{align}
Here, $K_0$ is the normalization factor defined in \cref{eq:ck20_coeffs_2} with $f_{0w}$ replaced by $f_{0}^{\rm L.O.}$. The force $F_z$ in the source term for the CP-odd transport equation should be the CP-violating part, $F_z=F_{o}\approx s\left(|m|^2\theta^\prime\right)^\prime/(2E_0 E_{z})$ (ignoring the second term in \cref{eq:Fz_decomp} contributed by $\Delta E$).
One can check that indeed these L.O. coefficients are the low $v_w$ limits of the full coefficients given by \cref{eq:coefficient CK20}.
Since only the leading order of $v_w$ is kept, this framework is not expected to perform well at larger $v_w$. This expectation is borne out in the numerical results of the next subsection.

%%%%%%%%%%%%%%%%%%%%%%%%%%%%%%%%%%%%%%%%%%%%%%%%%%%%%%%%%
\subsection{Baryogenesis in the DarkCPV Model}\label{sec:DarkCPV}

EWBG in the DarkCPV model was discussed in Ref.~\cite{Carena:2018cjh,Carena:2019xrr,Carena:2022qpf}.
At the time of EWPT, the chiral fermion $\chi$ gains a spatially varying complex mass via coupling to the complex singlet scalar field profile as,
\begin{align}
    M_\chi = m_0 + |\lambda_c| \exp(i [\theta + \mathrm{arg}(\SC)]) |\SC| \,.
\end{align}
This complex mass breaks CP symmetry, hence the chiral fermions $\chi^{\mathstrut}_{L},\bar{\chi}^{\mathstrut}_L$ (and their charge conjugates $\bar{\chi}^{\mathstrut}_R,\chi^{\mathstrut}_R$) scatter off the bubble wall at different rates,
creating a chiral asymmetry.
To transfer this chiral asymmetry to the SM, we gauge the lepton number to $U(1)_l$ with the gauge boson $Z'$, and assign lepton numbers to $\chi$ and $\SC$, such that $\chi$ and the SM leptons are connected via the $Z'$ portal.
The transferred chiral asymmetry in SM leptons will be turned into a net baryon asymmetry by the sphaleron.

The chiral asymmetry carried by $\chi$ is quantified by the CP-odd chemical potential $\mu_{o}(z)=\mu_{\chi_L}-\mu_{\bar{\chi}_L}=-(\mu_{\chi_R}-\mu_{\bar{\chi}_R})$,  where the second equality follows from charge conjugation symmetry. The profile of $\mu_{o}(z)$ can be solved from Eq.~\eqref{eq:transport} with the CP-odd source terms\footnote{The relaxation method behaves the best in solving Eq.~\eqref{eq:transport} numerically and thus is adopted here, as discussed in Ref.~\cite{Cline:2020jre}. Shooting method, Runge-Kutta, or BDF all tend to be unstable.}.
In our model, the dark fermion only directly couples to the complex scalar $\SC$ (generating the mass $M_\chi$) and the gauge boson $Z'$. The gauge coupling is strongly constrained by the experimental bounds from LEP and dark matter direct detection (DD), and is negligible compared to the Yukawa coupling. Therefore, we will consider the chirality flipping term generated by the mass $M_\chi$ to be the only collision term in the transport equation for $\chi^{\mathstrut}_{L,R}$, which can be computed following~\cite{Lee:2004we}.
Here, we take the approximation provided in~\cite{Kamada:2016eeb} and write the collision term as\footnote{Note that the collision term provided in those references is defined for the Boltzmann equation in terms of the particle number density $n$, which is widely used in the VIA formalism. Eq.~\eqref{eq:collision} is for the Boltzmann equation in terms of chemical potential. The relation between them is reviewed in Ref.~\cite{Cline:2021dkf}.}:
\begin{align}
    \label{eq:collision}
    \Gamma_m(z) \simeq \frac{3}{\pi^2} \frac{|M_\chi(z)|^2}{T^2} \gamma_{\rm th},
\end{align}
where $\gamma_{\rm th}$ is the thermal width\footnote{This is also known as the damping rate of quasiparticles in the thermal plasma, which basically refers to the rate of the decoherence of thermally dressed particles. This thermal effect played an important role in EWBG in the minimal SM, see~\cite{Farrar:1993hn,Farrar:1993sp,Huet:1994jb,Gavela:1993ts,Gavela:1994dt,Gavela:1994yf}.} of the fermion in the plasma, and is computed as~\cite{Braaten:1992gd,Joyce:1994zn,Thoma:1994yw}:
\begin{align}
    \gamma_{\rm th} \simeq
    \frac{1}{16}\alpha_\lambda T\,,
\end{align}
where $\alpha_\lambda = |\lambda_c|^2/(4\pi)$.
Here we again omit the gauge boson contribution to the thermal width.

The chiral asymmetry computed above is transferred to the SM
as follows.
The chiral asymmetry of $\chi$ implies a net $U(1)_l$ charge density determined by the different charges carried by the left and right-handed components,
\begin{align}
    \rho_l(z) = \frac{1}{3} N_g T_n^2 \mu_{o}(z).
\end{align}
Next, a long-range Coulomb-like potential of the $Z'$ can be generated by this net charge, i.e.
\begin{align}
    \langle Z'_0(z)  \rangle = \frac{g'}{2 M_{Z'}} \int_{-\infty}^{\infty} dz' \rho_l(z') \exp(- M_{Z'} |z - z'|).
\end{align}
This potential acts effectively as a chemical potential $\Delta\mu=g'\langle Z'_0(z)  \rangle$, sourcing a chiral asymmetry in the SM lepton sector with its equilibrium value given by,
\begin{align}
    \Delta n_{L_L}^{EQ}(z) = \frac{2 N_g T_n^2}{3} g'\langle Z_0'(z)\rangle.
\end{align}
This asymmetry biases the electroweak sphaleron to generate net lepton number, and thus net baryon number as,
\begin{align}
    \Delta n_B =\Delta n_{L_L}= \frac{\Gamma_{\rm sph}}{v_w \gamma_w} \int_0^\infty dz\, \Delta n_{L_L}^{EQ}(z) \exp(- \frac{\Gamma_{\rm sph}z}{v_w \gamma_w}),
\end{align}
with $\Gamma_{\rm sph}\simeq 10^{-6} T_n$. This baryon number behaves as $g^2/M_{Z'}^2$. Note that the $\gamma_w$ factor appearing in this expression is the result of boosting the wall frame back into the cosmic frame, as shown in Ref.~\cite{Cline:2020jre}. This factor is ignored in methods taking the L.O. small $v_w$ limit.

\begin{figure}[t]
    \centering
    \includegraphics[width=0.8\linewidth]{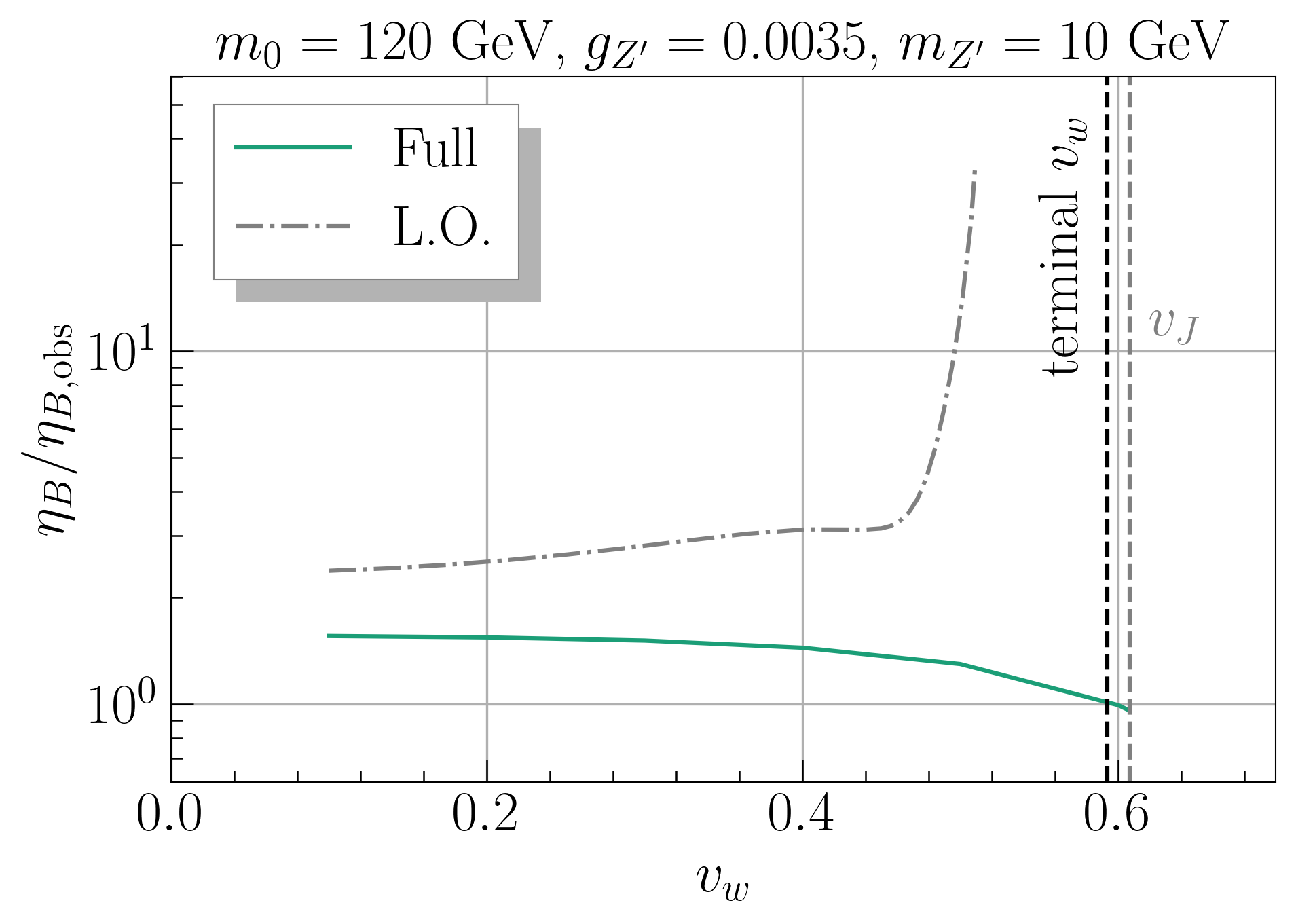}
    \caption{Baryon number to photon ratio $\eta_B$ normalized to the observed value $\eta_{B, \text{obs}}$ as a function of $v_w$ in the DarkCPV model. The terminal $v_w$ obtained using the entropy conservation approach is labeled by the black dashed line. The gray dashed line denotes the Jouguet velocity. The green solid curve shows the result of the all orders calculation, retaining full $v_w$-dependence. The gray dot-dashed line is the result of the L.O. approximation, which truncates the transport equation at leading order in $v_w$ and so is valid only for small $v_w \ll 1$. This approximation introduces an unphysical divergence when $v_w\sim 0.5$ crosses the sound speed. Clearly, the all orders treatment is necessary in such a regime.}
    \label{fig:bau_vw}
\end{figure}

Fig.~\ref{fig:bau_vw} shows the produced baryon number asymmetry normalized to the
observed value $\eta_{B, \rm obs} \equiv n_{B, \rm obs}/s\simeq 8.7 \times 10^{-11}$~\cite{Planck:2018vyg} as a function of $v_w$ calculated with the full and L.O. frameworks.
The benchmark point for this plot is chosen to be $m_a=10\,$GeV, $m_s=70\,$GeV, $m_0=120\,$GeV, $|\lambda_c|=0.3$, $\theta=-0.25$, $v_S' = 1048.04\,\gev$, $\lambda_{SH} = 9.38 \times 10^{-3}$, which leads to a hybrid bubble wall. We choose $g_{Z'} = 0.0035$ to avoid the LEP bound for $Z'$ boson search~\cite{Carena:2004xs,Jeong:2015bbi,Carena:2018cjh,Carena:2019xrr,Carena:2022qpf} as well as to ensure the BAU at the terminal $v_w$ assumes the observed value. The terminal velocity obtained from the entropy conservation approach is $v_w\approx 0.6$ as shown by the black dashed line in the plot.

Meanwhile, the BAU is not plotted for detonation bubble walls ($v_w > v_J$) in Fig.~\ref{fig:bau_vw} for the following reasons.
First, by definition, detonation has all the fluid perturbations falling behind the wall, as a result, the functions modeling the departure from equilibrium may be discontinuous at the wall, which was not carefully modeled in our calculational framework.
Second, detonation bubble walls may collide with each other before reaching the ultra-relativistic regime (i.e. percolation may occur before the end of acceleration).
Current understanding within this community for percolation is based on a static wall velocity~\cite{Ellis:2018mja}.
This may have significant impacts on the produced baryon number.
Based on these two points, theories for bubble percolation and baryogenesis for accelerating detonation walls remain as open questions.

Fig.~\ref{fig:bau_vw} has profound implications for the impact of $v_w$ on the computation of the BAU.
In the range near the terminal $v_w$, some early methods keeping only the leading-order terms of $v_w$ cannot accurately compute BAU.
We show the result of the L.O. approximation of Ref.~\cite{Cline:2000nw} as an example, which features a singularity at $v_w$ around 0.5, similarly to Ref.~\cite{Fromme:2006wx}. Clearly for such velocities, one requires a method which keeps the full $v_w$ dependence. Our all orders treatment is shown to behave smoothly in the deflagration/hybrid regime. Furthermore, due to the qualitative difference between deflagration/hybrid and detonation bubbles for BAU calculations as discussed in the previous paragraph, it is of vital importance to distinguish the fluid profiles from the first principles.

The benchmark point chosen in Fig.~\ref{fig:bau_vw} passes all bounds except those from the direct direction (DD) of DM-nucleon scattering~\cite{LZ:2024zvo}.
The relevant scattering cross section for DD is generated at the one-loop level and can be found in Ref.~\cite{Carena:2018cjh,Carena:2022qpf}.
These can easily be circumvented in a few different ways.
One option is to allow the $Z'$ boson to have a kinetic mixing with the photon.
Such a mixing does not alter the generated baryon number~\cite{Carena:2018cjh,Carena:2019xrr}, but it produces an additional diagram contributing to the DM-nucleon scattering cross section.
By properly choosing the sign and size of the mixing parameter, the new diagram can cancel the loop-level diagrams so that the DD bound is avoided.
For the benchmark used in Fig.~\ref{fig:bau_vw}, the required size of mixing is roughly $\epsilon e \sim 4 \times 10^{-4}$.
Another option is to introduce soft $U(1)$ breaking Majorana mass terms $\delta_L \chi_L \chi_L$ and $\delta_R \chi_R \chi_R$ with $\delta_{L,R} \ll m_0$.
The mass eigenstates then have a small mass splitting $\Delta m \simeq (\delta_L + \delta_R)/2$ and their vector current coupling with the $Z'$ boson becomes off-diagonal.
The vector-current DM-nucleon cross section is thus inelastic and kinematically suppressed by the mass splitting, and DD bound can be avoided for $\Delta m \simeq O(200~\rm keV)$~\cite{Krall:2017xij,Eby:2019mgs}.
For original references of this idea, see Ref.~\cite{Tucker-Smith:2001myb,Hall:1997ah}.
The axial-vector current, on the other hand, provides a suppressed cross section in the non-relativistic limit.
See, for example, Ref.~\cite{Kumar:2013iva}.

%%%%%%%%%%%%%%%%%%%%%%%%%%%%%%%%%%%%%%%%%%%%%%%%%%%%%%%%%
%%%%%%%%%%%%%%%%%%%%%%%%%%%%%%%%%%%%%%%%%%%%%%%%%%%%%%%%%
\section{Consequences for Gravitational Wave Signal}\label{sec:GWsignal}

First-order phase transitions proceed via bubble nucleation and source tensor perturbations\footnote{Tensor perturbations produced in a strong first-order phase transition may also lead to observable B-mode polarization in the CMB, particularly for late-time or supercooled transitions~\cite{Greene:2024xgq}. This can offer a complimentary signature of this scenario.} --- and therefore gravitational waves (GWs) --- through bubble collisions, sound waves, and turbulence~\cite{Kamionkowski:1993fg,Caprini:2015zlo,Weir:2017wfa}. During the initial bubble collision stage, mergers of true vacuum bubbles break spherical symmetry, allowing the gradient energy of the scalar field to source GWs~\cite{Hawking:1982ga,Kosowsky:1992vn}. This stage completes quickly, but can be the dominant GW source for strong, vacuum energy-dominated transitions, which are typically associated with significant supercooling and ``runaway'' bubble walls.

Long after the initial bubble collisions have completed,
shells of fluid kinetic energy continue to propagate through the plasma and collide, acting as an additional, much longer-lasting source of GWs. Fluid perturbations can be decomposed into longitudinal (compressional) and transverse (rotational) components, which source sound waves and magnetohydrodynamic (MHD) turbulence, respectively \cite{Hindmarsh:2015qta,Hindmarsh:2016lnk,Hindmarsh:2017gnf,Caprini:2006,Caprini:2009yp,Auclair:2022jod}.
Numerical simulations reveal that for intermediate strength phase transitions, compressional waves dominate over rotational waves \cite{Hindmarsh:2013xza}. Thus, sound waves are usually considered to be the dominant source of GWs for thermal transitions, and have been extensively studied for various models~\cite{Baldes:2018nel,Hall:2019ank,Hall:2019rld,Carena:2019une,Xie:2020bkl,Xie:2020wzn,Fujikura:2021abj,Cline:2021iff}.

It is convenient to decompose the energy density in GWs, as quantified by the spectral density parameter
\begin{equation}
    \Omega_{\rm GW} \equiv \frac{1}{\rho_{\rm tot}} \frac{d \rho_{\rm GW}}{d \ln k} \,,
\end{equation}
into contributions from each of these three sources
\begin{equation}
    \Omega_{\rm GW} = \Omega_{\phi} + \Omega_{\rm sw} + \Omega_{\rm tur} \,,
\end{equation}
with $\Omega_{\phi}$ corresponding to bubble collisions, $\Omega_{\rm sw}$ to sound waves, and $\Omega_{\rm tur}$ to MHD turbulence\footnote{It has recently been shown that particle production from bubble-bubble collision can act as an additional source of GWs~\cite{Inomata:2024rkt}. This source is non-negligible only in vacuum or supercooled phase transitions, similarly to the scalar field contribution.}. As previously alluded to, the relative contribution from each of these components depends sensitively on the transition strength, as quantified by the parameter $\alpha_n$ defined in Eq.~\eqref{eq:generalalpha}, as well as the bubble wall velocity $v_w$. In the following subsections, we detail how to calculate the GW power spectrum from each source, highlighting the dependence on $v_w$. Using this machinery, in Sec.~\ref{sec:GWplot} we show sample GW spectra for the DarkCPV model.

%%%%%%%%%%%%%%%%%%%%%%%%%%%%%%%%%%%%%%%%%%%%%%%%%%%%%%%%%
\subsection{Scalar Field Contribution}
\label{sec:scalar GW}

When a bubble nucleates, a portion of the vacuum energy is transformed into the gradient energy of the scalar whose interpolating profile defines the bubble wall. Once spherical symmetry is broken by bubble collision, this gradient energy can source GWs. For details, we refer the reader to Ref.~\cite{Jinno:2016vai}, which computes the GW spectrum semi-analytically from first principles by evaluating the 2-point function of the scalar stress-energy tensor.
The resulting GW spectrum takes the form
\begin{align}\label{eq:GW collision}
    \Omega_{\phi}h^2 & = 1.67 \times 10^{-5} \left( \frac{\kappa_\phi\, \alpha_n}{1 + \alpha_n} \right)^2 \left( \frac{H}{\beta} \right)^{2} \left( \frac{100}{g_*} \right)^{1/3} \Delta_{\rm peak} S
\end{align}
with coefficients
\begin{align}
     & \Delta_{\rm peak} =\left( \frac{0.48 v_w^3}{1 + 5.3 v_w^2 + 5 v_w^4} \right)\,,\nonumber                                                                                                               \\
     & S = \left( 0.064 \left( \frac{f}{f_{\rm env}} \right)^{-3} + (1-0.064-0.48) \left( \frac{f}{f_{\rm env}} \right)^{-1} + 0.48 \left( \frac{f}{f_{\rm env}} \right)\right)^{-1}\,, \nonumber             \\
     & f_{\rm env} = 16.5 \times10^{-9} \mathrm{Hz}\left( \frac{0.35}{1 + 0.069 v_w + 0.69 v_w^4} \right) \left( \frac{\beta}{H} \right) \left( \frac{T}{100} \right) \left( \frac{g_*}{100} \right)^{1/6}\,.
\end{align}
Here, $\beta/H$ describes the ratio of the Hubble timescale $H^{-1}$ to the duration of the phase transition $\beta^{-1}$ and can be evaluated as
\begin{align}
    \frac{\beta}{H} = T\frac{d(S_3/T)}{dT} \bigg|_{T_n}\,,
\end{align}
where $S_3$ is the three-dimensional Euclidean action. The coefficient $\kappa_\phi$ characterizes the fraction of released energy which goes into bubble gradient energy, and is believed to behave as~\cite{Weir:2017wfa,Ai:2024ikj}
\begin{align}
    \kappa_\phi = \frac{E_{\rm wall}}{E_{\rm vac}} \propto \frac{\gamma_w}{R}
\end{align}
where $R$ is the bubble radius at the time of bubble collision, and $E_{\rm wall}$ and $E_{\rm vac}$ represent the energy carried by the bubble wall and released from the phase transition, respectively.\footnote{In the presence of the plasma, $E_{\rm vac}$ may be replaced by a more proper quantity that reflects the total released energy, but the dependency on $R$ does not change.} As the bubble wall expands, this fraction is suppressed by $R$, making the produced GW signal negligible unless the wall continues to accelerate, allowing an increasing $\gamma_w$ to compensate for the growing $R$.

Whether or not the bubble wall can keep accelerating depends on its interactions with the plasma. It was once believed that once $\alpha_n$ exceeded some critical value, the bubble wall would accelerate without bound, becoming a ``runaway'' bubble~\cite{Bodeker:2009qy}. Working under the LTE approximation, detonation solutions naively seem to result in runaway bubbles, since there is no friction force to stop the bubble from accelerating. It has recently been realized, however, that an additional source of frictive pressure in the form of the emission of soft radiation as particles cross the bubble wall becomes important in the ultra-relativistic regime, growing with increasing $\gamma_w$~\cite{Bodeker:2017cim}.
Whether this pressure behaves as $\mathcal{P} \propto \gamma_w^2$~\cite{Hoche:2020ysm} or $\mathcal{P} \propto \gamma_w$~\cite{Gouttenoire:2021kjv} remains under debate.
Nevertheless, the consensus is that the pressure grows sufficiently quickly as the wall accelerates into the ultra-relativistic regime that it will ultimately reach a terminal\footnote{In the case of detonations, it is also possible that bubble walls may collide with one another before reaching the terminal velocity. To our knowledge, this scenario has not been investigated in the literature, though some papers have provided preliminary estimates~\cite{Ai:2024ikj}.} velocity. That is, so long as a thermal plasma is present, no runaway bubble wall can appear, and thus the GW contribution from the bubble collision stage is negligible. Only for strongly supercooled phase transitions, for which expansion significantly dilutes the plasma, can this source dominate. Since this is not the case for the DarkCPV model in our parameter space of interest, we omit the bubble collision contribution to the total GW signal in Fig.~\ref{fig:GW_dcpv}.

%%%%%%%%%%%%%%%%%%%%%%%%%%%%%%%%%%%%%%%%%%%%%%%%%%%%%%%%%
\subsection{Sound Waves}

Long after the initial bubble collisions have completed, fluid sound shells (``sound waves'') continue to collide and merge. This provides an additional source of GWs, which is actually the dominant contribution for most thermal transitions. The GW spectrum from this source takes the form~\cite{Hindmarsh:2013xza,Hindmarsh:2015qta,Caprini:2015zlo,Hindmarsh:2016lnk,Hindmarsh:2017gnf,Weir:2017wfa,Ellis:2020awk,Caprini:2024hue}
\begin{align}
    \Omega_{\rm sw}h^2  = 2.65 \times 10^{-6} \left( \frac{\kappa_{\rm sw} \, \alpha_n}{1 + \alpha_n} \right)^2 \left( \frac{H}{\beta} \right) \left( \frac{100}{g_*} \right)^{1/3} (H \tau_{\rm sw}) \, v_w \left( \frac{f}{f_{\rm sw}} \right)^3 \left( \frac{7}{4 + 3(f/f_{\rm sw})^2} \right)^{7/2} \,,
\end{align}
the sound wave peak frequency $f_{\rm sw}$ set by
\begin{align}
    \label{eq:fsw}
    f_{\rm sw} = 1.9 \times 10^{-5} \, \mathrm{Hz} \left( \frac{1}{v_w} \right) \left( \frac{\beta}{H} \right) \left( \frac{g_*(T_n)}{100} \right)^{1/6} \left( \frac{T_n}{100~\rm GeV} \right) \,.
\end{align}
The ``time duration suppression factor'' $H \tau_{\rm sw}$, describing the ratio of the sound wave formation timescale $\tau_{\rm sw}$ to the expansion timescale $H^{-1}$, has been applied in various approximations in the recent literature~\cite{Weir:2017wfa,Ellis:2018mja,Cutting:2019zws,Ellis:2020awk,Fujikura:2021abj,Cline:2021iff,Caprini:2024hue}.
Here we choose the approximation in Ref.~\cite{Ellis:2018mja,Fujikura:2021abj,Caprini:2024hue},
\begin{align}
    H \tau_{\rm sw} = \min \left( 1, (8 \pi)^{1/3} \left( \frac{\max (v_w, c_s)}{\beta/H} \right)\left( \frac{4}{3} \frac{1+\alpha_n}{\kappa_{\rm sw}\alpha_n} \right)^{1/2}  \right) \,.
\end{align}
The fraction of released energy which goes into sound waves, $\kappa_{\rm sw}$, is in principle calculable from the fluid velocity profile $v(\xi)$. Ref.~\cite{Espinosa:2010hh} provides a good numerical fit as a function of $\alpha_n$ in their Appendix A, which we reproduce here:
\begin{align}
    \kappa_{v_w \lesssim c_s} = & \frac{c_s^{11/5} \kappa_A \kappa_B}{(c_s^{11/5} - v_w^{11/5}) \kappa_B + v_w c_s^{6/5} \kappa_A}, \nonumber                                         \\
    \kappa_{c_s < v_w < v_J} =  & \kappa_B + (v_w - c_s) \delta \kappa + \frac{(v_w - c_s)^3}{(v_J - c_s)^3}\left( \kappa_C - \kappa_B - (v_J - c_s) \delta \kappa \right), \nonumber \\
    \kappa_{v_w \to 1} =        & \frac{\alpha_n}{0.73 + 0.083 \sqrt{\alpha_n} + \alpha_n},
    \label{eq:kappa_sw}
\end{align}
where
\begin{align}
    \kappa_A & = v_w^{6/5} \frac{6.9 \alpha_n}{1.36 - 0.037 \sqrt{\alpha_n} + \alpha_n} \,, \,\,\, \kappa_B      = \frac{\alpha_n^{2/5}}{0.017 + (0.997 + \alpha_n)^{2/5}}, \nonumber \\
    \kappa_C & = \frac{\sqrt{\alpha_n}}{0.135 + \sqrt{0.98 + \alpha_n}} \,, \,\,\,
    \delta \kappa = -0.9 \log \frac{\sqrt{\alpha_n}}{1 + \sqrt{\alpha_n}} \,.
\end{align}
We comment that the above numerical fit was based on the range $0.001 < \alpha_n < 10$, which is consistent with the range of $\alpha_n$ explored in our current work.

%%%%%%%%%%%%%%%%%%%%%%%%%%%%%%%%%%%%%%%%%%%%%%%%%%%%%%%%%
\subsection{Magnetohydrodynamic Turbulence}

If the timescale $\tau_{\rm sw}$ is shorter than the Hubble time, $H \tau_{\rm sw} < 1$, MHD turbulence can form~\cite{Hindmarsh:2013xza,Hindmarsh:2015qta,Hindmarsh:2016lnk,Hindmarsh:2017gnf}. In general, the contribution from MHD must be determined through numerical simulations~\cite{RoperPol:2019wvy}. From modeling Kolmogorov-type turbulence \cite{Caprini:2009yp,Caprini:2015zlo,Weir:2017wfa}, though, it is expected that the parameter dependency can be expressed as
\begin{align}
    \Omega_{\rm tur}h^2 = 3.35 \times 10^{-4} \left( \frac{\kappa_{\rm turb} \, \alpha_n}{1 + \alpha_n} \right)^{3/2} \left( \frac{H}{\beta} \right) \left( \frac{100}{g_*} \right)^{1/3} v_w S_{\rm turb} \,,
\end{align}
where
\begin{align}
    \label{eq:turb}
     & S_{\rm turb} = \frac{(f/f_{\rm turb})^3}{(1 + (f/f_{\rm turb}))^{11/3} (1+8\pi f/h_*)} \,, \nonumber                                                                                                        \\
     & f_{\rm turb} = 2.7 \times 10^{-5} \, \mathrm{Hz} \left( \frac{1}{v_w} \right) \left( \frac{\beta}{H} \right) \left( \frac{g_*(T_n)}{100} \right)^{1/6} \left( \frac{T_n}{100~\rm GeV} \right) \,, \nonumber \\
     & h_* = 16.5 \times 10^{-6} \, \mathrm{Hz} \left( \frac{g_*(T_n)}{100} \right)^{1/6} \left( \frac{T_n}{100~\rm GeV} \right) \,.
\end{align}
The coefficient $\kappa_{\rm tur}$ describes the fraction of vacuum energy that is transformed into MHD turbulence, and is estimated to be between $(0.05-0.1)\kappa_{\rm sw}$ for thermal phase transitions. We assume $\kappa_{\rm turb} = 0.05 \kappa_{\rm sw}$ in this work.

%%%%%%%%%%%%%%%%%%%%%%%%%%%%%%%%%%%%%%%%%%%%%%%%%%%%%%%%%
\subsection{GW Spectra in the DarkCPV Model}\label{sec:GWplot}

\begin{figure}[t]
    \centering
    \includegraphics[width=0.8\linewidth]{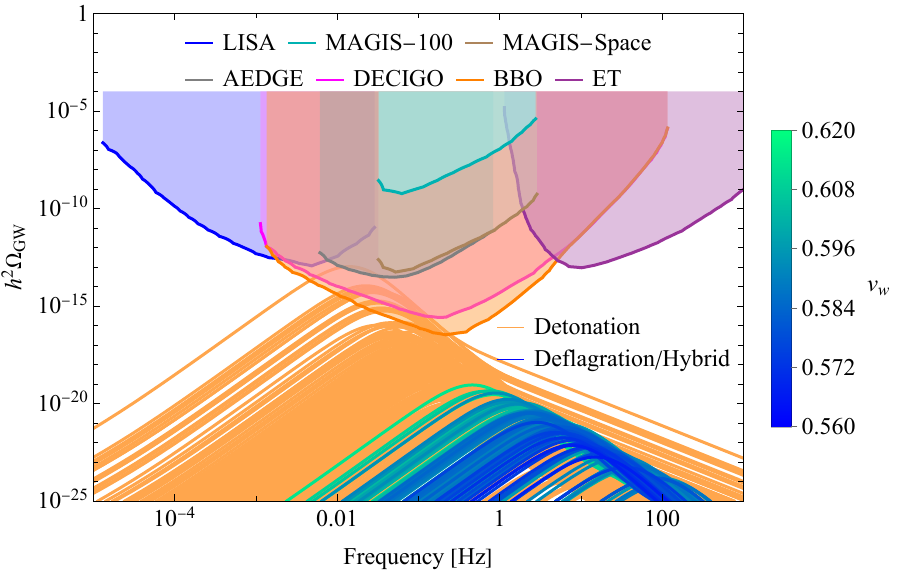}
    \caption{GW spectrum for the DarkCPV model, with $v_w$ depicted by the color gradient. Orange curves are detonation bubbles.}
    \label{fig:GW_dcpv}
\end{figure}

In Fig.~\ref{fig:GW_dcpv}, we show the GW spectrum for the DarkCPV model using the data in Fig.~\ref{fig:dcpv_scatter}. For the deflagration/hybrid walls, the true $v_w$ as shown in \fig{dcpv_scatter}{c} is used, while for the detonation walls, we take $v_w=1$. We observe a positive correlation between the amplitude of GW signals and $v_w$ --- as one might expect, the detonation walls generally produce stronger GW signals than the deflagration/hybrid walls, and deflagration/hybrid walls with higher $v_w$ generally produce stronger GW signals than those with lower $v_w$. We also observe a negative correlation between the amplitude and the peak frequency of the spectrum, reflecting the negative correlation of the peak frequency with $v_w$, as can be seen from \cref{eq:fsw,eq:turb}.

Deflagration/hybrid bubbles in the DarkCPV model occur only for very small $\alpha_n\lesssim 4\times 10^{-3}$, as shown in the upper right panel of \cref{fig:dcpv_scatter}. As a consequence of these small $\alpha_n$ and the fact that $\Omega_{\rm GW} \sim (\frac{\alpha_n}{1+\alpha_n})^2$, the GW signal from the deflagration/hybrid transitions of this model are below the sensitivities of all proposed experiments shown in \cref{fig:GW_dcpv}.
Thus, the observation of a  GW signature, if interpreted in the context of the DarkCPV model, would be indicative of a detonation solution and therefore potentially incompatible with a successful baryogenesis in this model.
%endanger the explanation of baryogenesis within this model.
We emphasize that the GW spectra from first-order phase transitions are highly model-dependent, since the upper bound of $\alpha_n$ permitting deflagration/hybrid solutions depends sensitively on the particle content and model details, as discussed in \cref{sec:vw computation}.
The situation may be more optimistic for other models for which the deflagration/hybrid solution class admits larger $\alpha_n$, such that a GW signal would be compatible with EWBG.

%%%%%%%%%%%%%%%%%%%%%%%%%%%%%%%%%%%%%%%%%%%%%%%%%%%%%%%%%
%%%%%%%%%%%%%%%%%%%%%%%%%%%%%%%%%%%%%%%%%%%%%%%%%%%%%%%%%
\section{Conclusion}\label{sec:conclusion}

In this paper, we have carefully examined the bubble wall dynamics during a first-order EWPT and calculated from first principles the terminal wall velocity $v_w$. After verifying that the dominant contribution to the hydrodynamic obstruction indeed comes from the equilibrium distribution function, we work in the local thermal equilibrium (LTE) approximation. In this regime, one may use entropy conservation to provide an additional matching condition, allowing for a greatly simplified computation of the bubble wall velocity. We have introduced efficient algorithms for calculating $v_w$ for each fluid profile solution class --- deflagration, hybrid, and detonation.

The above has allowed us to perform a thorough examination of the parameter dependencies of $v_w$ in two exemplary models capable of a strong first-order EWPT --- a $\mathbb{Z}_2$-symmetric real singlet extension of the SM, and the ``DarkCPV'' model of Refs.~\cite{Carena:2018cjh,Carena:2019xrr,Carena:2022qpf}. In particular, we studied correlations between $v_w$ and two common measures of the phase transition strength, $\alpha_n$ and $v_n/T_n$. Interestingly, in both models we find $v_w$ and $v_n/T_n$ to be well modeled by functions of the form $f(x) = a \ln(x) + b + c \, x + d \, x^2$, with $x=\alpha_n$ and $\{ a, b, c, d\}$ fitting coefficients. The coefficients for $v_n/T_n$ are quite similar for both models, while those for $v_w$ differ greatly between models (and depend on which parameters are held fixed). This potentially indicates that the relationship between $v_n/T_n$ and $\alpha_n$ may be less model dependent, though further work is needed to confirm this.

For the DarkCPV model, we also studied the implications of $v_w$ for baryogenesis and the GW signal. We reviewed the calculation of the BAU in different computational frameworks, commenting on their applicability given the value of $v_w$. In particular, the treatment based on truncating the transport equations at leading order in $v_w$ was determined to be unreliable for realistic values of $v_w$ in this model. This indicates that one should first carefully determine $v_w$ from the first principles before choosing a BAU computation framework. The GW signal is another observable which depends sensitively on both the wall velocity $v_w$ and the transition strength $\alpha_n$. We found that in the DarkCPV model, deflagration/hybrid solutions occur only for very small $\alpha_n\lesssim 4\times 10^{-3}$. Consequently, the GW signal, which scales as $(\frac{\alpha_n}{1+\alpha_n})^2$, is beyond the sensitivities of current and projected GW detectors for this class of solutions. Detonation solutions, on the other hand, may yield a detectable GW signal, but would not necessarily be compatible with a successful baryogenesis.

Finally, we emphasize that the bubble wall dynamics are only well-understood under LTE or near LTE conditions, which provides a reliable framework for the deflagration/hybrid profiles. For detonation bubble walls, instead, out-of-equilibrium contributions become important. Bubble collision and merging may also occur during the bubble acceleration period, such that no terminal velocity is ever reached. The dynamics of accelerating bubble walls remains an open area of research in this community and requires further exploration.

\acknowledgments
We are especially grateful to Wen-Yuan Ai for insightful discussions during the completion of this work.
We also would like to thank James Cline, Shang Ren, Peizhi Du, Zhen Liu, Yikun Wang, Yue Zhang, Gordan Krnjaic, Keisuke Harigaya, and Vincenzo Cirigliano for many useful discussions.
IRW and TO are grateful to Perimeter Institute of Theoretical Physics for hosting and support during part of the work.
MC research at Perimeter Institute is supported in part by the Government of Canada through the Department of Innovation, Science and Economic Development, and by the Province of Ontario through the Ministry of Colleges and Universities.
AI is supported by NSF Grant PHY-2310429, Simons Investigator Award No.~824870, DOE HEP QuantISED award \#100495, the Gordon and Betty Moore Foundation Grant GBMF7946, and the U.S.~Department of Energy (DOE), Office of Science, National Quantum Information Science Research Centers, Superconducting Quantum Materials and Systems Center (SQMS) under contract No.~DEAC02-07CH11359.
IRW is supported by the DOE distinguished scientist fellowship grant FNAL 22-33.
TO is supported by the Visiting Scholars Program of URA at Fermilab and the DOE Office of Science Distinguished Scientist Fellows Award 2022.
Fermilab was operated by Fermi Research Alliance, LLC under Contract No.~DE-AC02-07CH11359 with the U.S. Department of Energy, Office of Science, Office of High Energy Physics, and is now operated by Fermi Forward Discovery Group, LLC under Contract No. 89243024CSC000002 with the U.S. Department of Energy, Office of Science, Office of High Energy Physics.

%%%%%%%%%%%%%%%%%%%%%%%%%%%%%%%%%%%%%%%%%%%%%%%%%%%%%%%%%
%%%%%%%%%%%%%%%%%%%%%%%%%%%%%%%%%%%%%%%%%%%%%%%%%%%%%%%%%
\appendix
\setcounter{equation}{0}

\section{Out-of-Equilibrium Bubble Wall}\label{app:non_LTE}

This section reviews the bubble wall dynamics for a general out-of-equilibrium fluid, complementing the treatment in Sec.~\ref{sec:bubblewallreview}. The starting point is again EMT conservation \cref{eq:conservation}, but with the more general expression for the fluid EMT given by \cref{eq:Tf}. Focusing on a single-component fluid for simplicity, we decompose the distribution function into equilibrium and out-of-equilibrium contributions: 
\begin{equation}
    f(k,z) = f^{\rm eq}(k,z) +  f^{\rm out}(k,z) \,.
\end{equation}
Correspondingly, the fluid EMT can also be decomposed as 
\begin{equation}
    T^{\mu \nu}_f = T_{f, \, \rm LTE}^{\mu \nu}+T_{f, \, \rm out}^{\mu \nu} \,,
\end{equation}
where
\begin{equation}
    T_{f,\,\text{out}}^{\mu \nu} = \int \frac{d^3k}{(2\pi)^3} \frac{k^{\mu}k^{\nu}}{2E_k} f^{\rm out}(k,z) \,.
\end{equation}
Recall that the equilibrium contribution to $T^{\mu \nu}_f$ could equivalently be expressed in the perfect fluid form of Eq.~(\ref{eq:perfectfluid}). Similarly, one can always perform a decomposition of $T^{\mu \nu}_{f\, \text{out}}$ into a linear combination of Lorentz covariant pieces, 
\begin{equation}
    T_{f,\,\text{out}}^{\mu \nu} = T_{f,\,\text{out}}^{(\eta)} \eta^{\mu \nu} + T_{f,\,\text{out}}^{(u)} u^\mu u^\nu + T_{f,\,\text{out}}^{(\bar{u})} \bar{u}^\mu \bar{u}^\nu + T_{f,\,\text{out}}^{(u\bar{u})} (u^\mu \bar{u}^\nu + \bar{u}^{\mu} u^\nu ) \,,
\end{equation}
where 
\begin{equation}\label{eq:Tfouteta}
    T_{f,\,\text{out}}^{(\eta)} = \int \frac{d^3 k}{(2\pi)^3} \frac{1}{2E_k} \left( m^2 + (k_\mu \bar{u}^\mu)^2 - (k_\mu u^\mu)^2 \right)  f^{\rm out}(k,z) \,.
\end{equation}
The explicit forms of $T_{f,\,\text{out}}^{(u)}$, $T_{f,\,\text{out}}^{(\bar{u})}$, and $T_{f,\,\text{out}}^{(u\bar{u})}$ will not be important for what follows, and can be found in Ref.~\cite{Laurent:2022jrs}. 

The equilibrium contribution to the divergence, $\partial_\mu T_{f, \rm LTE}^{\mu \nu}$, was derived in \cref{sec:bubblewallreview} and is given by \cref{eq:partial_Tf_LTE}. It can be combined with the scalar field divergence of Eq.~(\ref{eq:divTphi}) and the out-of-equilibrium contribution to give the following conservation law:
\begin{equation}\label{eq:EMT_non_LTE}
\begin{split}
    \partial^\nu \phi \bigg( \Box \phi + & \frac{\partial V_{\rm eff}}{\partial \phi} + \frac{\partial T_{f,\,\text{out}}^{(\eta)}}{\partial \phi} \bigg) + \partial_\mu (w u^\mu u^\nu) - s \partial^\nu T \\
    & + \partial_\mu \left( T_{f,\,\text{out}}^{(u)} u^\mu u^\nu + T_{f,\,\text{out}}^{(\bar{u})} \bar{u}^\mu \bar{u}^\nu + T_{f,\,\text{out}}^{(u\bar{u})} (u^\mu \bar{u}^\nu + \bar{u}^{\mu} u^\nu ) \right) = 0 \,,
\end{split}
\end{equation}
The quantity in parentheses must vanish independently, giving the following effective EOM for the scalar field
\begin{equation}
    \Box \phi + \frac{\partial V_{\rm eff}}{\partial \phi} + \frac{\partial T_{f,\,\text{out}}^{(\eta)}}{\partial \phi} = 0 \,.
\end{equation}
Noting that only the term $\propto m^2$ in Eq.~(\ref{eq:Tfouteta}) is $\phi$-dependent, this simplifies to
\begin{equation}\label{eq:phieom}
    \Box \phi + \frac{\partial V_{\rm eff}}{\partial \phi} + \frac{\partial m^2}{\partial \phi} \int \frac{d^3 k}{(2\pi)^3} \frac{1}{2E_k}  f^{\rm out}(k,z) = 0 \,.
\end{equation}

\subsection{Dynamics of the Expanding Bubble}
\label{sec:expansion}
Consider now an isolated, spherically symmetric bubble of true vacuum that has nucleated in the false vacuum and is expanding radially outwards, as in Fig.~\ref{fig:coordinate}. This expansion is the result of a competition between the driving pressure from the energy released in the phase transition and backreaction from the plasma in the form of a frictive pressure on the wall. If the friction exerted by the plasma on the wall is sufficient to balance the driving force, the wall quickly reaches the terminal velocity $v_w$. If not, the bubble wall continues to accelerate towards ultra-relativistic velocities.

In order to work out expressions that will allow us to discriminate between the two cases, we work in the \textit{instantaneous}\footnote{Because we allow the acceleration to be non-zero here, we cannot define a global wall rest frame. Nevertheless, we can still define an \textit{instantaneous} wall rest frame.} wall rest frame, in which all the quantities are functions of $z$. Our starting point is the scalar equation of motion of Eq.~(\ref{eq:phieom}) (generalized to multiple species). Multiplying both sides by $(\partial_z \phi)$ and integrating over the wall, we find
\begin{equation}\label{eq:balance}
    \int dz \, (\partial_z \phi) \left( \ddot{\phi} + \frac{\partial V_0}{\partial \phi} + \sum_i n_i \frac{dm_i^2}{d\phi} \int \frac{d^3k}{(2\pi)^3} \frac{1}{2 E_i} f_i(k,z) \right) = 0 \,,
\end{equation}
where we have used the fact that $\int dz \, (\partial_z \phi) (\partial_z^2 \phi) = \frac{1}{2}\int dz \, \partial_z \left[ (\partial_z \phi)^2 \right] = 0$ is a total derivative and $\partial_z \phi$ vanishes away from the wall in order to eliminate the would-be second term. The left-most term is related to the wall's acceleration and may be identified with the net pressure, which we define as
\begin{equation}\label{eq:Pnet}
\begin{split}
    \mathcal{P}_{\rm net} & \equiv \int dz (\partial_z \phi) \ddot{\phi}\\
    & = - \int dz (\partial_z \phi) \left( \frac{\partial V_0}{\partial \phi} + \sum_i n_i \frac{dm_i^2}{d\phi} \int \frac{d^3k}{(2\pi)^3} \frac{1}{2 E_i} f_i(k,x) \right) \,.
\end{split}
\end{equation}
The condition for a stationary wall is that the net pressure vanishes, $\mathcal{P}_{\rm net} = 0$. Note that we follow the sign convention of Ref.~\cite{Konstandin:2014zta,Laurent:2022jrs,Ai:2024shx}, for which a negative $\mathcal{P}_{\rm net}$ is associated with an accelerating wall.

It is instructive to decompose $\mathcal{P}_{\rm net}$ as the difference between friction and driving pressure, 
\begin{equation}
    \mathcal{P}_{\rm net} = \mathcal{P}_{\rm fric} - \mathcal{P}_{\rm drive} \,.
\end{equation}
We identify\footnote{Note that this decomposition is not unique, and various authors may have different conventions.} the driving pressure with the difference in the zero-temperature potential between the symmetric and broken phases, 
\begin{equation}
    \mathcal{P}_{\rm drive} = \int dz \, (\partial_z \phi) \frac{\partial V_0}{\partial \phi} = \Delta V_0 \,,
\end{equation}
which is purely determined by the zero-temperature scalar potential.
The remaining term is then identified as the friction caused by the plasma,
\begin{equation}\label{eq:Pfric}
\begin{split}
    \mathcal{P}_{\rm fric} & = - \int dz \, (\partial_z \phi) \sum_i n_i \frac{dm_i^2}{d\phi} \int \frac{d^3k}{(2\pi)^3} \frac{1}{2 E_k} f_i(k,x).
\end{split}
\end{equation}
A negative $\mathcal{P}_{\rm net}$ implies $\mathcal{P}_{\rm drive} > \mathcal{P}_{\rm fric}$, corresponding to an accelerating wall. Typically, this acceleration continues until the pressures balance and $\mathcal{P}_{\rm net}=0$, at which point the wall assumes steady state motion at the terminal velocity $v_w$. 
As we will see in the following subsection, the friction, and thus the net pressure, peaks at the Jouguet velocity $v_J$.
If the friction at this peak is not strong enough to balance the driving pressure, $\mathcal{P}_{\rm fric}< \mathcal{P}_{\rm drive}$ at $v_J$, the wall will accelerate beyond $v_J$ and enter the detonation regime.
In this regime, the LTE contribution to the friction is negligible, since the plasma is at rest in front of the bubble wall. The bubble wall will keep accelerating to the ultra-relativistic regime, where the pressure receives further contributions from the emission of soft radiation, and will finally reach a terminal velocity~\cite{Bodeker:2017cim,Hoche:2020ysm,Gouttenoire:2021kjv}.

One may observe from Eq.~(\ref{eq:Pfric}) that $\mathcal{P}_{\rm fric}$ receives contributions only from those particles with $\phi$-dependent masses coupling directly to the wall. This, however, is not the full picture, given that the distribution function for a given species depends on interactions with \textit{all} the particles in the plasma. Even if a particle does not couple directly to $\phi$, it may still contribute to $\mathcal{P}_{\rm fric}$ provided that it interacts sufficiently strongly with a particle which does couple directly to the wall.
Following Ref.~\cite{Ai:2024shx}, we refer to those particles that couple directly to the wall and have $\phi$-dependent masses  ``active'' particles, and those that couple indirectly through their interactions with active particles ``passive'' particles. Active particles always contribute to the effective degrees of freedom of the plasma, while passive particles should only be counted if their interactions are sufficiently rapid that the active particles can communicate to them the influence of the wall on a time scale much faster than the phase transition itself. A natural condition for passive particles to count towards the degrees of freedom is then
\begin{equation}\label{eq:condition}
    \Gamma = n \expval{\sigma v} \gg \beta \,,
\end{equation}
where $\Gamma$ is the interaction rate for the relevant scatterings that help to maintain the passive species in kinetic equilibrium with the plasma and $\beta$ is the inverse duration of the phase transition.
Passive particles that fail to satisfy this criterion are essentially transparent to the wall.
Note that all SM particles are either ``active'' or have a large enough interaction rate with the wall, and so their particle degrees of freedom contribute to $V_{\rm eff}$.

\subsection{Wall Velocity and Thickness from EOM}
\label{app:vw_Lwall_CL}
The bubble wall velocity $v_w$ enters the expression of the net pressure \cref{eq:Pnet} as a parameter, and can be solved for by requiring $\mathcal{P}_{\rm net}=0$. In addition to $v_w$, $\mathcal{P}_{\rm net}$ also depends on the scalar field profile across the space, which can be parameterized by, e.g., the wall width $L_\phi$. More equations are needed to solve for the field profile parameters, which can be constructed as moment equations of the scalar field EOM \cref{eq:phieom}. A general out-of-equilibrium calculation is performed in \cite{Laurent:2022jrs} for the first-order EWPT occurring from the $\mathbb{Z}_2$-symmetric real singlet extension model.
In the following, we will review the calculations in \cite{Laurent:2022jrs} in the LTE approximation, and compare them to the entropy conservation approach discussed in \cref{sec:vwinLTE}.
Readers are referred to the original work for the case including out-of-equilibrium contribution.
This section adopts the $\mathbb{Z}_2$-symmetric singlet scalar extension model described in Sec.~\ref{sec:Z2_model} as an example, and the discussions can be easily generalized to more complicated models.

In such a model model, both $h$ and $S$ obey analogous EOMs, given by Eq.~(\ref{eq:phieom}). For convenience, we define the left-hand side of these equations as $E_{h,S}$. In the wall frame and assuming LTE, these are
\begin{subequations}
\begin{equation}
    E_h \equiv - \partial_z^2 h + \frac{\partial V_{\rm eff}}{\partial h} = 0 \,,
\end{equation}
\begin{equation}
    E_S \equiv - \partial_z^2 S + \frac{\partial V_{\rm eff}}{\partial S} = 0 \,,
\end{equation}
\end{subequations}
where the finite-temperature effective potential $V_{\rm eff}(h,S; T)$ receives contributions from both $h$ and $S$. A simple Ansatz for the field profiles solving these equations is
\begin{subequations}\label{eq:scalar profile}
\begin{equation}
    h(z) = \frac{h_0}{2} \left( 1 - \tanh \left( \frac{z}{L_h} \right) \right) \,,
\end{equation}
\begin{equation}
    S(z) = \frac{S_0}{2} \left( 1 + \tanh \left( \frac{z}{L_S} + \delta_S \right) \right) \,,
\end{equation}
\end{subequations}
where $L_h$ and $L_S$ are the wall widths and $\delta_S$ is the offset between the two wall centers, all of which are constants once the walls have reached the terminal velocity. The vacuum expectation values $h_0$ and $S_0$, on the other hand, are generically functions of the asymptotic plasma temperatures $T_\pm$, which in turn depend on $v_w$. 

The temperature itself also smoothly changes from $T_-$ to $T_+$ across $-\infty < z < \infty$. 
The EMT conservation along $z$ can be used to determine the temperature profile $T(z)$.
In the wall frame, it can be written as (under LTE):
\begin{align}
\label{eq:EMT along wall}
    T^{30} &= w \gamma^2 v = c_1 \nonumber \\
    T^{33} &= \frac{1}{2} \left((\partial_z h)^2 + (\partial_z S)^2\right) - V_{\rm eff}(h(z), S(z), T(z)) + w \gamma^2 v^2 = c_2\,,
\end{align}
where $c_1$ and $c_2$ are constants that can be directly determined from $v_\pm$ and $T_\pm$.
Combining these two equations to eliminate $v$, one arrives at
\begin{align}
\label{eq:solve T}
    \frac{1}{2} \left((\partial_z h)^2 + (\partial_z S)^2\right) - V_{\rm eff}(h(z), S(z), T(z)) - \frac{1}{2}w + \frac{1}{2} \sqrt{4 c_1^2 + w^2} - c_2 = 0\,.
\end{align}
With this equation, the temperature $T(z)$ can be solved directly, using the scalar field profile given in Eq.~\eqref{eq:scalar profile}.

All the above discussions depend on parameters $L_h$, $L_s$ and $\delta s$.
We can solve for the values of these parameters by taking moments of $E_h$ and $E_S$ and demanding that they vanish. Two convenient choices are the pressures $\mathcal{P}_{h,S}$ and pressure gradients $\mathcal{G}_{h,S}$ associated to each field~\cite{Konstandin:2014zta,Laurent:2022jrs}. As in Eq.~(\ref{eq:Pnet}), we can define the net pressures on the walls
\begin{subequations}\label{eq:Pdefs}
\begin{equation}
    \mathcal{P}_h(v_w, L_h, L_S, \delta_S) = - \int dz (\partial_z h) E_h \,,
\end{equation}
\begin{equation}
    \mathcal{P}_S(v_w, L_h, L_S, \delta_S) = - \int dz (\partial_z S) E_S \,.
\end{equation}
\end{subequations}
These quantities are well-suited to determine $v_w$, since the walls reach their steady state when the net pressures on each of them vanish, $\mathcal{P}_{h,\,S}=0$. 
Additionally, the requirement $\mathcal{P}_h = \mathcal{P}_S$ ensures constant offset between the walls, and can be used to solve for $\delta_S$. The net pressure gradients on the walls are
\begin{subequations}\label{eq:Gdefs}
\begin{equation}
    \mathcal{G}_h(v_w, L_h, L_S, \delta_S) = \int dz\, (\partial_z h) \left( \frac{2 h}{h_0} - 1 \right) E_h \,,
\end{equation}
\begin{equation}
    \mathcal{G}_S(v_w, L_h, L_S, \delta_S) = \int dz\, (\partial_z S) \left( \frac{2 S}{S_0} - 1 \right) E_S \,.
\end{equation}
\end{subequations}
A non-vanishing pressure gradient $\mathcal{G}_{h,S}$ would lead to the compression or stretching of the wall, and so demanding that these moments vanish allows us to determine the constant wall widths $L_{h,S}$. A final steady state solution thus requires 
\begin{equation}
    (\mathcal{P}_h + \mathcal{P}_S) = (\mathcal{P}_h - \mathcal{P}_S) = \mathcal{G}_h = \mathcal{G}_S = 0 \,.
\end{equation}

We can use these conditions to solve for the wall velocity $v_w$ following the algorithm of CL.
Since the framework to find out $v_J$, solving for $v_\pm$ and $T_\pm$, and iterate over $v_w$ is essentially the same as that in Sec.~\ref{sec:vwinLTE},
we only present how to compute $P_{\rm tot}$ for a given $v_w$.
The algorithm is
\begin{enumerate}
    \item Determining $c_{1,2}$ with $T_\pm$ and $v_\pm$ from Eq.~\eqref{eq:EMT along wall}\footnote{Theoretically, input $T_-$ and $v_-$ should give the same result as $T_+$ and $v_+$. However, we found it numerically more stable to start from $v_-$ and $T_-$ and solve for $T(z)$ from $-\infty$ to $+\infty$.}.
    \item Design a function to solve for $T(z)$ as a function of $L_h$, $L_s$, $\delta s$, using the $c_{1,2}$ and Eq.~\eqref{eq:solve T}.
    \item Given $T(z)$, compute $\mathcal{P}_{h,S}$ and $\mathcal{G}_{h,S}$ as in Eqs.~(\ref{eq:Pdefs}) and (\ref{eq:Gdefs}) as a function of $L_h$, $L_s$, $\delta s$.
    \item Demand $(\mathcal{P}_h - \mathcal{P}_S) = \mathcal{G}_h = \mathcal{G}_S = 0$ and solve for $L_h$, $L_S$, and $\delta_S$. 
    \item Calculate $\mathcal{P}_{\rm tot}(v_w)$ using the solutions for $\{L_h, L_S, \delta_S \}$.
\end{enumerate}

\begin{figure}[h]
    \centering
    \includegraphics[width=0.45\linewidth]{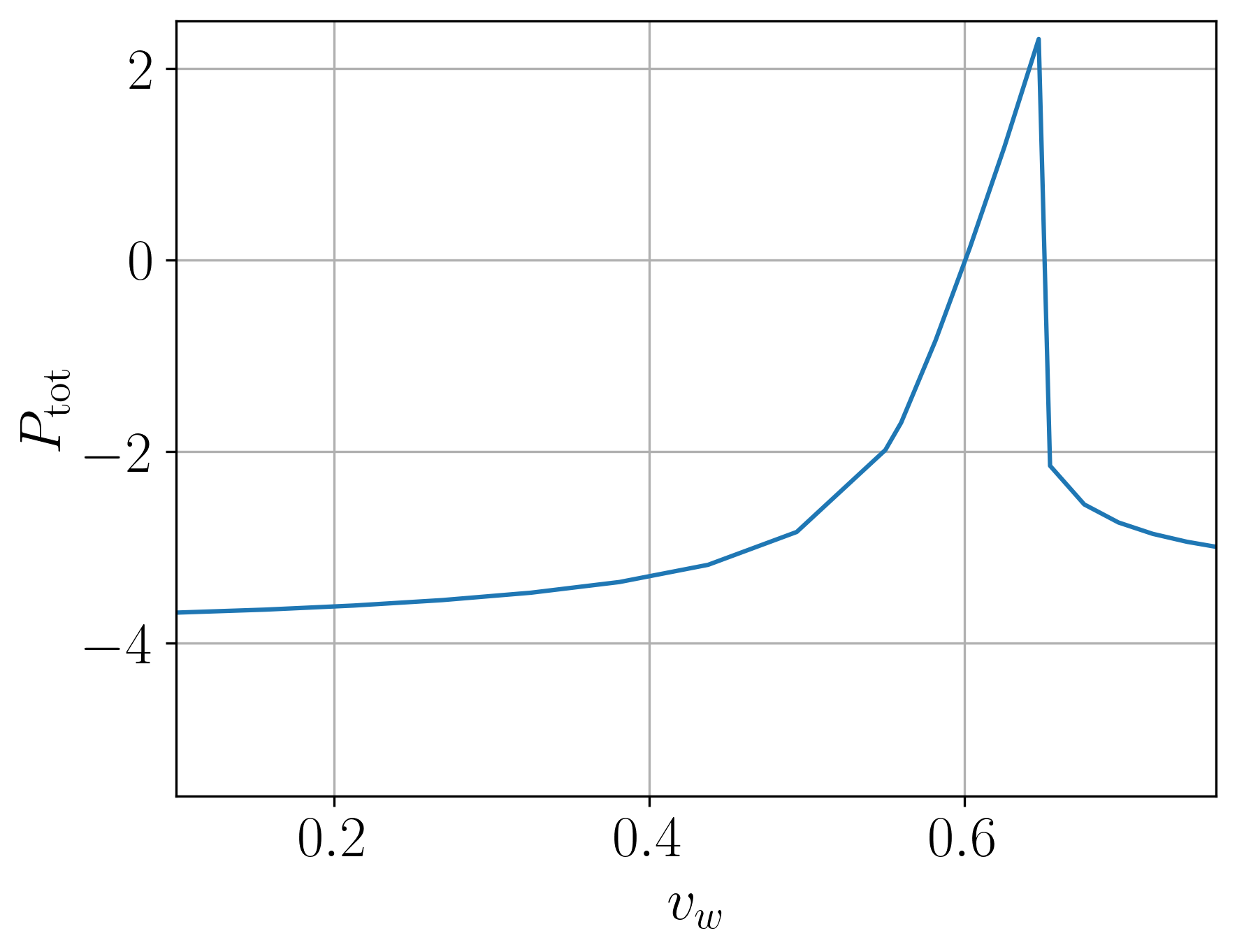}
    \includegraphics[width=0.45\linewidth]{Figure/ds.png}
    \caption{Left: Total pressure on the wall $\mathcal{P}_{\rm tot}$ as a function of $v_w$ in arbitrary units. Right: Entropy difference $\Delta(s \gamma v)$ between plasmas on either side of the wall (arbitrary units). Note that both display qualitatively identical behavior and cross zero at the same value of $v_w$.
    We do not plot $v_w$ up to 1 since LTE is not justified in that range.}
    \label{fig:peakplot}
\end{figure}

\begin{table}[h]
    \centering
    \begin{tabular}{|c||c|c|}
        \hline
         & scalar EOM & Entropy conservation \\
        \hline\hline
          Condition 1, 2  & \multicolumn{2}{c|}{Matching conditions between $v_{+}$ and $v_-$ (\cref{eq:mc1} and \cref{eq:mc2})} \\
          \hline
        Condition 3 & Scalar field EOM Eq.~\eqref{eq:phieom} & Entropy conservation Eq.~\eqref{eq:mc3} \\
        \hline
        Advantages & \makecell{Can go beyond LTE\\Can output wall width} & \makecell{Fast\\Numerically stable} \\
        \hline
        Disadvantages & \makecell{Slow\\Numerical instabilities} & \makecell{Cannot go beyond LTE\\Cannot compute wall widths}\\
        \hline
    \end{tabular}
    \caption{Summary of different algorithms to solve for the wall velocity.}
    \label{tab:solution of vw}
\end{table}
The left panel of Fig.~\ref{fig:peakplot} shows the $P_{\rm tot}$ as a function of $v_w$.
The benchmark chosen for this plot is $m_S = 66.14~\gev$, $\lambda_{SH} = 0.79$, $\lambda_S = 1$.
The friction opposing bubble expansion comes mainly from the heated plasma in front of the wall --- a phenomenon known as ``hydrodynamic obstruction'' \cite{Konstandin:2010dm,Laurent:2022jrs}. This is a purely equilibrium effect which appears whenever there is a shock wave preceding the wall (i.e., in the deflagration and hybrid regimes). As one increases $v_w$, the total pressure also begins increasing from its initial negative value. (Recall that we have defined a negative $\mathcal{P}$ to correspond to an accelerating wall.) After passing the Jougeut velocity $v_J$, $\mathcal{P}_{\rm tot}$ drops sharply since the heated plasma front disappears. Thus, there should be a sharp peak in $\mathcal{P}_{\rm tot}$ at $v_w = v_J$, as is indeed the case in Fig.~\ref{fig:peakplot}.

The right panel of Fig.~\ref{fig:peakplot} shows the $\Delta (s\gamma v)$ using the same parameter point (which is the same plot in Fig.~\ref{fig:ds_vw}) for the comparison between the approaches of the scalar EOM and entropy conservation.
We observe that the curves in these two panels have a similar shape and reach zero at an identical $v_w$,
indicating the equivalence between the two approaches.
The entropy conservation approach significantly reduces computation time and minimizes numerical instability, as it eliminates the need to solve for $T(z)$ and $(L_h, L_S, \delta_S)$. For models with additional scalar fields, requiring even more free parameters to solve the wall profile, the entropy conservation method is even more advantageous.
We summarize the scalar EOM and entropy conservation approaches in Table~\ref{tab:solution of vw}.

\section{Free Energy}\label{app:freeenergy}

The central quantity governing the thermodynamics of the phase transition is the free energy density $\mathcal{F}$, which recieves contributions from both the scalar field and fluid $\mathcal{F} = \mathcal{F}_\phi + \mathcal{F}_f$.
\begin{comment}
which is equivalent to the finite temperature contribution to the effective potential $\mathcal{F}_f = V_T$ and sets the fluid pressure as
\begin{equation}
    p_f = - \mathcal{F}_f = - V_T \,.
\end{equation}
\end{comment}
The free energy of the fluid is equivalent to the finite temperature contribution to the effective potential $\mathcal{F}_f = V_T$ and is computed as the sum of the free energies of the thermalized particle constituents,
\begin{equation}
    \mathcal{F}_f = \sum_B n_B \mathcal{F}_B + \sum_F n_F \mathcal{F}_F \,,
\end{equation}
with $n_{B/F}$ the number of degrees of freedom per species.
In order for a particle to count towards the degrees of freedom, it should either couple directly to the wall or interact sufficiently strongly with particles coupling directly to the wall.
See Eq.~(\ref{eq:condition}) of Appendix~\ref{sec:expansion} for a precise formulation of this condition and related discussion.
We remark that all SM particles trivially satisfy this condition at the electroweak scale.
To compute this explicitly, recall that the free energy density of a single bosonic degree of freedom with field-dependent effective mass $m$ is
\begin{equation}
\label{eq:FB}
    \mathcal{F}_B = T \int \frac{d^3 k}{(2\pi)^3} \ln \left( 1 - e^{-E_k/T} \right) \equiv T^4 J_B\left(\frac{m}{T} \right) \,,
\end{equation}
where we have defined the bosonic thermal function $J_B$. Similarly for a fermionic degree of freedom,
\begin{equation}
\label{eq:FF}
    \mathcal{F}_F = - T \int \frac{d^3 k}{(2\pi)^3} \ln \left( 1 + e^{-E_k/T} \right) \equiv T^4 J_F\left(\frac{m}{T} \right) \,,
\end{equation}
with $J_F$ the fermionic thermal function. These admit the following high temperature expansions
\begin{subequations}
\label{eq:free_energy_FB}
    \begin{equation}
        \mathcal{F}_B \simeq - \frac{\pi^2}{90} T^4 + \frac{1}{24} m^2 T^2 - \frac{1}{12\pi} m^3 T - \frac{1}{32\pi^2} m^4 \left[ \ln \left( \frac{m e^{\gamma_E}}{4\pi T} \right) - \frac{3}{4} \right] \,,
    \end{equation}
    \begin{equation}
        \mathcal{F}_F \simeq - \frac{7}{8} \frac{\pi^2}{90} T^4 + \frac{1}{48} m^2 T^2 - \frac{1}{32\pi^2} m^4 \left[ \ln \left( \frac{m e^{\gamma_E}}{\pi T} \right) - \frac{3}{4} \right] \,.
    \end{equation}
\end{subequations}
The free energy in the high temperature approximation can then be expressed as
\begin{equation}\label{eq:Ffluid}
    \mathcal{F}_f = - \frac{\pi^2}{90} g_\star T^4 + \frac{T^2}{24} \left( \sum n_B m_B^2 + \frac{1}{2} \sum_F n_F m_F^2 \right) - \frac{T}{12 \pi} \sum n_B m_B^3 \,,
\end{equation}
where $g_\star = \sum_B n_B + \frac{7}{8} \sum_F n_F$ is the effective number of degrees of freedom and we have neglected higher order terms for simplicity. The free energy density for the scalar field, meanwhile, is simply the zero-temperature potential\footnote{The zero-temperature potential includes both the tree-level piece and the zero-temperature loop-level corrections, $V_0 = V_{\rm tree} + V_0^{\rm loop}$.}
\begin{equation}
    \mathcal{F}_\phi = V_0 \,.
\end{equation}
The total free energy density for the scalar-fluid system is then the finite-temperature effective potential
\begin{equation}\label{eq:Ftotal}
    \mathcal{F} = V_0 + V_T \equiv V_{\rm eff} \,,
\end{equation}
which takes the schematic form
\begin{equation}\label{eq:Veffgenform}
    V_{\rm eff} = V_0 - \frac{\pi^2}{90} g_\star T^4 + \frac{T^2}{24} \left( \sum n_B m_B^2 + \frac{1}{2} \sum_F n_F m_F^2 \right) - \frac{T}{12 \pi} \sum n_B m_B^3 + ... 
\end{equation}

Accurate predictions of phase transition parameters therefore depend on an accurate determination of $V_{\rm eff}$.
For practical reasons, the finite-temperature effective potential is often computed in perturbation theory, though there are a number of theoretical uncertainties associated with a perturbative calculation \cite{Croon:2020cgk,Chala:2024xll,Chala:2025aiz}.
Among them, the finite temperature perturbative expansion breaks down at high temperatures when infrared bosonic modes become highly occupied.
This generically necessitates the resummation of large thermal corrections to the effective potential \cite{Parwani:1991gq, Arnold:1992rz,Bahl:2024ykv,Bittar:2025lcr}.

\section{Reference Frames}\label{app:frames}

This appendix section provides a more detailed explanation of the two main reference frames that are used in Sec.~\ref{sec:bubblewallreview}.
Namely, we will consider the rest frame of the bubble center, which we will refer to as the ``cosmic frame'', and the rest frame of the bubble wall, which we will call the ``wall frame''. Both frames are depicted in Fig.~\ref{fig:coordinate}.

\subsubsection*{Cosmic Frame}
\label{sec:frames}

In the cosmic frame, the spherical symmetry and lack of characteristic scale motivate us to introduce the dimensionless coordinate $\xi \equiv R/t$, where $R$ is the distance as measured from the center of the bubble and $t$ is the time since nucleation. A fluid element at the point described by $\xi$ in the wave profile has velocity $v(\xi)$, and so in the cosmic frame (in spherical coordinates) we parameterize the fluid 4-velocity $u_{\rm cf}^\mu$ and the 4-velocity orthogonal to the flow $\bar{u}_{\rm cf}^\mu$ as
\begin{equation}\label{eq:ucf}
    u_{\rm cf}^\mu(\xi) = \gamma_{\rm cf}(\xi) \big(1, 0,0,v_{\rm cf}(\xi) \big) \,, \,\,\,\, \bar{u}_{\rm cf}^\mu(\xi) = \gamma_{\rm cf}(\xi) \big( v_{\rm cf}(\xi), 0,0,1\big) \,,
\end{equation}
where $\gamma(\xi) = 1/\sqrt{1 - v_{\rm cf}(\xi)^2}$ is the Lorentz factor and $v_{\rm cf}(\xi)$ is the fluid velocity in the cosmic frame. 

\begin{figure}[t!]
\centering
\includegraphics[width=0.9\linewidth]{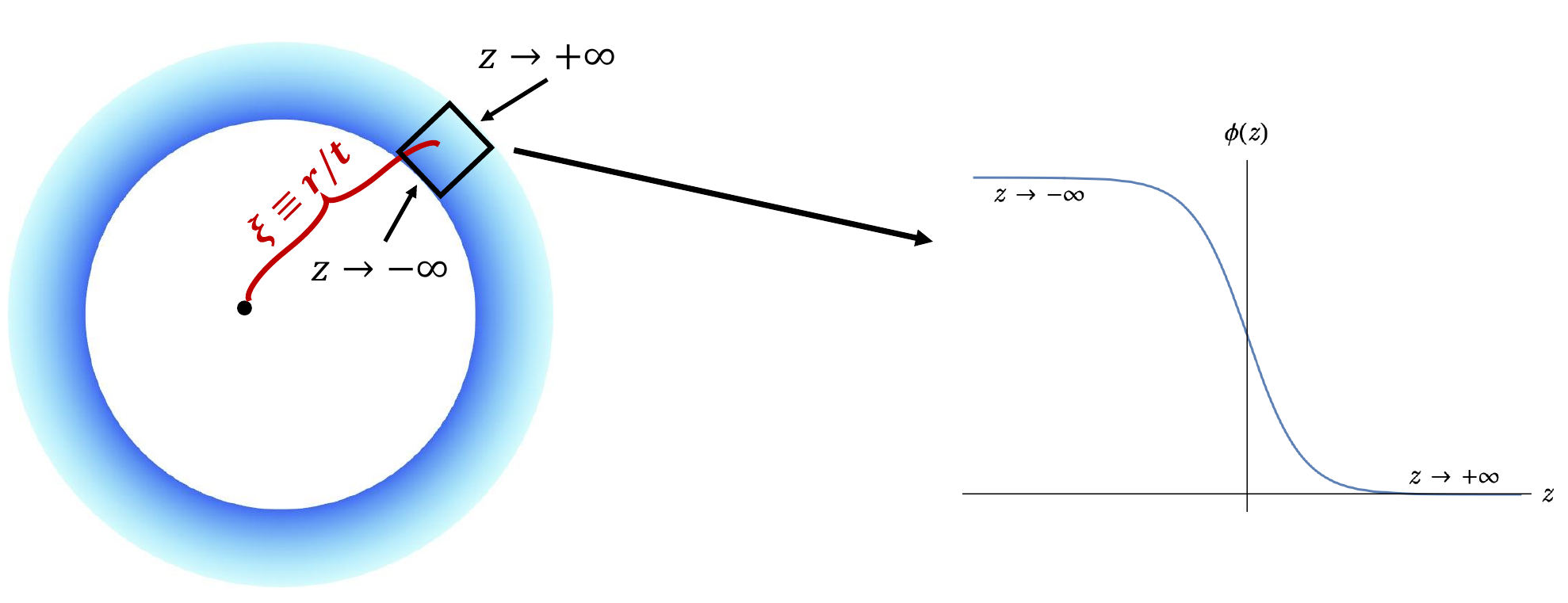}
\caption{The two main reference frames and their coordinate systems. The left panel shows a spherically symmetric bubble expanding radially outwards in the cosmic frame, with $\xi = r/t$ labeling an arbitrary point in the wave profile. The bubble wall itself is shown in blue; note the thickness has been greatly exaggerated for illustrative purposes. The gradient represents how the scalar profile interpolates between false and true vacua, as shown more explicitly in the right panel as a function of $z$, the wall frame coordinate. 
}\label{fig:coordinate}
\end{figure}

We presume that the wall quickly\footnote{Though the initial period of acceleration is short relative to the duration of the phase transition, it is nevertheless important, as demonstrated by the numerical simulations of \cite{Krajewski:2024gma}. It would be valuable to develop an analytic description of this period.} accelerates to its terminal velocity $v_w$, after which it undergoes steady-state expansion at $v_w$. The wall is then located at $\xi_w = v_w$ in this coordinate system, and we label the fluid outside and inside the wall by $\xi_+$ and $\xi_-$, respectively, with $\xi_+ > \xi_w$ and $\xi_- < \xi_w$. We denote the scalar field values on either side of the wall by $\phi_\pm$, which satisfy
\begin{equation}
    \frac{\partial V_{\rm eff}}{\partial \phi} \bigg|_{\phi_\pm, T_\pm} = 0 \,, \,\,\,\,\,\, V_{\rm eff}(\phi_-, T_-) < V_{\rm eff}(\phi_+, T_+) \,,
\end{equation}
consistent with the true, broken vacuum lying in the bubble interior and the false, symmetric vacuum outside. The bubble wall is the region interpolating between these two local minima, and is often modeled with a $\tanh$ profile; see Fig.~\ref{fig:coordinate}. The bubble wall width is typically very small compared with the radius, motivating us to work in the thin wall approximation while in the cosmic frame. 

\subsubsection*{Wall Frame}

In other contexts, it will be more convenient to work in the wall's rest frame, where all quantities are time-independent. It is always possible to define such a frame when the bubble wall moves with constant velocity $v_w$. A sufficiently large bubble can be approximated as locally planar, and without loss of generality, we take the wall to be at $z=0$ and consider expansion in the $+z$-direction from the perspective of the cosmic frame, such that the fluid velocity is in the $-z$-direction. The broken and symmetric phases are found in the limits $z \rightarrow -\infty$ and $z \rightarrow + \infty$, respectively; see Fig.~\ref{fig:coordinate}. We parameterize the fluid 4-velocity $u_{\rm wf}^\mu$ and the 4-velocity orthogonal to the flow $\bar{u}_{\rm wf}^\mu$ as
\begin{equation}
    u_{\rm wf}^\mu(z) = \gamma(z) \big( 1, 0, 0, - v_{\rm wf}(z) \big) \,, \,\,\,\, \bar{u}_{\rm wf}^\mu(z) = \gamma(z) \big( v_{\rm wf}(z), 0, 0, -1 \big) \,,
\end{equation}
where we have introduced the minus sign so that $v_{\rm wf} \geq 0$ is non-negative. For future reference, we also define the asymptotic values of the velocity in the symmetric and broken phases, $v_+ \equiv v_{\rm wf}(z \rightarrow + \infty)$ and $v_- \equiv v_{\rm wf}(z \rightarrow - \infty)$.

\section{Relativistic Fluid Equations}
\label{app:fluid}

The matching conditions derived in Sec.~\ref{sec:matching_conds} will turn out to be sufficient to solve for the bubble wall velocity in the case of detonation profiles (see Sec.~\ref{sec:type}). However, for deflagration and hybrid profiles --- which feature a shock front preceding the bubble wall --- we need a bit more information to fully characterize the relativistic velocity and temperature profiles. Again, our starting point will be energy-momentum conservation. 

The total energy-momentum tensor is given by the sum of Eqs.~(\ref{eq:Tphi}) and (\ref{eq:perfectfluid}). We work in the cosmic frame away from the wall, where the gradient energy of the scalar field vanishes. In this region, the conservation law $\partial_\mu T^{\mu \nu} = 0$ reads
\begin{equation}
    \partial_\mu (w u_{\rm cf}^\mu) u_{\rm cf}^\nu + w u_{\rm cf}^\mu (\partial_\mu u_{\rm cf}^\nu) - \partial^\nu p = 0 \,.
\end{equation}
Projecting along the flow by contracting both sides with the fluid 4-velocity $u^{\rm cf}_\nu$ defined in Eq.~(\ref{eq:ucf}), we find
\begin{equation}\label{eq:fluideq1}
    \partial_\mu (w u_{\rm cf}^\mu) = u^\nu_{\rm cf} \partial_\nu p \,,
\end{equation}
where we have used the property $u_\nu \partial_\mu u^\nu = 0$. If we instead project orthogonal to the flow by contracting with the space-like vector $\bar{u}_\nu^{\rm cf}$, we find
\begin{equation}\label{eq:fluideq2}
    w u_{\rm cf}^\mu \bar{u}_{\rm cf}^\nu \partial_\mu u_\nu^{\rm cf} = \bar{u}^\nu_{\rm cf} \partial_\nu p \,.
\end{equation}
Expressing the gradient contractions in terms of partial derivatives with respect to $\xi$ via $u_{\rm cf}^\mu \partial_\mu = - \frac{\gamma_{\rm cf}}{t} (\xi - v_{\rm cf}) \partial_\xi$ and $\bar{u}_{\rm cf}^\mu \partial_\mu = \frac{\gamma_{\rm cf}}{t} (1 - \xi v_{\rm cf} ) \partial_\xi$, Eqs.~(\ref{eq:fluideq1}) and (\ref{eq:fluideq2}) can be re-written as
\begin{subequations}\label{eq:relfluideqs}
    \begin{equation}
        \left( \frac{\xi - v_{\rm cf}}{w} \right) \frac{\partial e}{\partial \xi} = 2 \frac{v_{\rm cf}}{\xi} + \left( 1 - \gamma_{\rm cf}^2 v_{\rm cf} (\xi - v_{\rm cf}) \right) \frac{\partial v_{\rm cf}}{\partial \xi} \,,
    \end{equation}
    \begin{equation}
        \left( \frac{1 - v_{\rm cf} \, \xi}{w} \right) \frac{\partial p}{\partial \xi} = \gamma_{\rm cf}^2 (\xi - v_{\rm cf}) \frac{\partial v_{\rm cf}}{\partial \xi} \,.
    \end{equation}
\end{subequations}
Finally, these equations can condensed into a single expression by invoking the plasma sound speed $c_s^2$ defined in Eq.~(\ref{eq:soundspeed}).
Thus we arrive at~\cite{Espinosa:2010hh,Laurent:2022jrs}
\begin{subequations}
\begin{equation}\label{eq:masterfluidappendixversion}
    2 \frac{v_{\rm cf}(\xi)}{\xi} = \gamma_{\rm cf}^2 (1 - v_{\rm cf}(\xi) \, \xi) \left( \frac{\mathfrak{v}(\xi, v_{\rm cf}(\xi))^2}{c_s^2(\xi)} - 1 \right) \frac{\partial v_{\rm cf}(\xi)}{\partial \xi} \,,
\end{equation}
\begin{equation}\label{eq:mastertemp}
    \frac{\partial T}{\partial \xi} = T \gamma_{\rm cf}^2 \mathfrak{v}(\xi, v_{\rm cf}) \frac{\partial v_{\rm cf}}{\partial \xi} \,.
\end{equation}
\end{subequations}
Specifically, Eq.~\eqref{eq:masterfluidappendixversion} is nothing but Eq.~\eqref{eq:masterfluid}.
These are the master equations for the relativistic velocity
and temperature profiles, respectively, and generically need to be solved numerically.
The numerical solution is typically easier with recasting these equations in parametric form~\cite{Espinosa:2010hh}
\begin{equation}
\begin{split}
    \frac{d v}{d \tau} & = 2 v c_s^2 (1-v^2)(1-\xi v) \,, \\
    \frac{d \xi}{d \tau} & = \xi \left( (\xi - v)^2 - c_s^2(1-\xi v) \right) \,, \\
    \frac{d T}{d \tau} & = \partial_\xi T \frac{d \xi}{d \tau} \,.
\end{split}
\end{equation}

\section{Effective Potential of Singlet Scalar Extension and DarkCPV Model}\label{app:singletscalar}

Two models are applied in the main text of the current work.
In both models,
we parameterize the Higgs doublet as 
\begin{equation}
    H = \frac{1}{\sqrt{2}} \begin{pmatrix} \chi_1 + i \chi_2 \\ h + i \chi_3 \end{pmatrix} \,,
\end{equation}
with $\chi_1, \chi_2, \chi_3$ Goldstone bosons and $h$ the Higgs boson.
In the unitary gauge, the tree-level potential of the singlet scalar with $\mathbb{Z}_2$ symmetry is then
\begin{equation}
    V_{\rm tree} = - \frac{\mu_H^2}{2} h^2 + \frac{\lambda_H}{4} h^4 + \frac{\mu_S^2}{2} S^2 + \frac{\lambda_S}{4} S^4 + \frac{\lambda_{SH}}{4} h^2 S^2 \,.
\end{equation}
In the DarkCPV model, the complex scalar is written as $\SC = s + i a$, such that
\begin{align}
    V_{\rm tree} = &- \frac{\mu_H^2}{2} h^2 + \frac{\lambda_H}{4} h^4 + \frac{\lambda_{\SC H}}{4} h^2(s^2 + a^2) \nonumber \\
    & - \frac{\mu_{\SC}^2}{2}(s^2 + a^2) + \frac{\lambda_{\SC}}{4} (s^2 + a^2)^2 + \kappa_{\SC}^2 (s^2 - a^2)\,,
\end{align}
where $\mu_{H(\SC)}^2\equiv \lambda_{H(\SC)} v_{H(\SC)}^2$.
The 1-loop effective potential has the following general form
\begin{equation}
    V_{\rm eff} = V_{\rm tree} + V_{\rm CW} + V_T^{1\text{-loop}} \,,
\end{equation} 
which receives both radiative and finite-temperature corrections, as captured by the Coleman-Weinberg potential $V_{\rm CW}$ and 1-loop thermal potential $V_T^{1\text{-loop}}$, respectively.
Working in the $\overline{\text{MS}}$ scheme with renormalization scale $\mu_R$, the Coleman-Weinberg potential is~\cite{ColemanWeinberg:1973}
\begin{equation}\label{eq:VCW}
    V_{\mathrm{CW}} = \frac{1}{64 \pi^2} \sum_i n_i \hat{m}_i^4 \left[ \ln \left( \frac{\hat{m}_i^2}{\mu_R^2} \right) - c_i \right] \,,
\end{equation}
where the index $i$ runs over the relevant species in the plasma with particle degrees of freedom\footnote{We work in the convention that fermions have negative particle degrees of freedom.} $n_i$, which in this case includes $h$, $\chi_{1,2,3}$, $t$, $W^\pm$, $Z$, and extra scalar $S$ (or $s$ and $a$ in DarkCPV).
\begin{align}
    n_S &= 1 \,, \,\,\, n_s = 1\,, \,\,\, n_a = 1\,, \,\,\, n_h = 1 \,, \,\,\, n_{\chi_{1,2,3}} = 1\,, \nonumber \\
    n_t &= -12 \,, \,\,\, n_W = 6 \,, \,\,\, n_Z = 3 \,.
\end{align}
The choice of renormalization scale $\mu_R$ is not unique.
In this work, we choose $\mu_R = m_t$ for both of the models.
The constant $c_i$ is equal to $3/2$ for scalars and fermions and equal to $5/6$ for the vector bosons, and the background field-dependent effective masses $\hat{m}$ for $\mathbb{Z}_2$ scalar model are
\begin{equation}
\begin{split}
    & \hat{m}_S^2 = \mu_S^2 + 3 \lambda_S S^2 + \frac{\lambda_{SH}}{2} h^2 \,, \\
    & \hat{m}_h^2 = - \mu_H^2 + 3 \lambda_H h^2 + \frac{\lambda_{SH}}{2} S^2 \,, \\
    & \hat{m}_{hS}^2 = \lambda_{SH} h S \,, \\
    & \hat{m}_{\chi_{1,2,3}}^2 = - \mu_H^2 + \lambda_H h^2 + \frac{\lambda_{SH}}{2} S^2 \,, \\
    & \hat{m}_t^2 = \frac{y_t^2}{2} h^2 \,, \,\,\, \hat{m}_W^2 = \frac{g^2}{4} h^2 \,, \,\,\, \hat{m}_Z^2 = \frac{{g'}^{\,2} + g^2}{4} h^2 \,.
\end{split}
\end{equation}
In the DarkCPV model, the gauge boson and top quark masses are the same as above, while the scalar masses are given by
\begin{align}
    &\hat{m}_{\chi_{1,2,3}}^2 = - \mu_H^2 + \lambda_H h^2 + \frac{\lambda_{\SC H}}{2}(s^2 + a^2)\,, \nonumber \\
    &\hat{m}_h^2 =- \mu_H^2 + 3\lambda_H h^2 + \frac{\lambda_{\SC H}}{2}(s^2 + a^2)\,, \nonumber \\
    &\hat{m}_s^2 = - \mu_\SC^2 + 3 \lambda_\SC s^2 + \lambda_\SC a^2 + \frac{\lambda_{\SC H}}{2}h^2 + 2 \kappa_\SC^2\,, \nonumber \\
    &\hat{m}_a^2 = - \mu_\SC^2 + 3 \lambda_\SC a^2 + \lambda_\SC S^2 + \frac{\lambda_{\SC H}}{2}h^2 + 2 \kappa_\SC^2\,, \nonumber \\
    &\hat{m}_{hs} = \lambda_{\SC H}s h\,, \nonumber \\
    &\hat{m}_{ha} = \lambda_{\SC H}a h\,, \nonumber \\
    &\hat{m}_{sa} = 2 \lambda_\SC s a\,,
\end{align}
Note that since the Higgs and scalar singlet mix, one must diagonalize in $h-S$ space or $h-s-a$ space to find the mass eigenvalues entering into $V_{\rm eff}$ above.
The 1-loop finite temperature potential is the same one as mentioned in Appendix~\ref{app:freeenergy}
\begin{equation}\label{eq:VT}
    V_{T}^{1\text{-loop}} = \frac{T^4}{2\pi^2} \sum_i n_i J_{B/F} \left( \frac{\hat{m}_i^2}{T^2} \right) \,, 
\end{equation}
where again $i$ runs over $\{ S (\text{or }s,a), h, \chi_{1,2,3}, t, W^\pm, Z \}$ we have defined the bosonic and fermionic thermal functions
\begin{equation}
    J_{B/F}(y^2) = \int_0^\infty dx\, x^2 \ln \left( 1 \mp e^{- \sqrt{x^2 + y^2}} \right) \,.
\end{equation}
In the high-temperature limit, these admit expansions
\begin{subequations}
\begin{equation}
    J_B(y^2 \ll 1) \simeq - \frac{\pi^4}{45} + \frac{\pi^2}{12} y^2 - \frac{\pi}{6} y^3 - \frac{1}{32} y^4 \ln \left( \frac{y^2}{a_B} \right) + ... \,,
\end{equation}
\begin{equation}
    J_F(y^2 \ll 1) \simeq \frac{7 \pi^4}{360} - \frac{\pi^2}{24} y^2 - \frac{1}{32} y^4 \ln \left( \frac{y^2}{a_F} \right) + ... \,,
\end{equation}
\end{subequations}
where $a_B = 16 \pi^2 \exp(\frac{3}{2}- 2 \gamma_E)$ and $a_F = \pi^2 \exp(\frac{3}{2} - 2 \gamma_E)$. The finite temperature part of the effective potential suffers from various issues in the IR as bosonic modes become highly occupied~\cite{Croon:2020,Bahl:2024ykv}. These issues can be alleviated by resumming the perturbative expansion in terms of a new convergent loop parameter. At 1-loop, it suffices to resum a class of diagrams called daisy diagrams, and replace the tree-level bosonic squared masses $\hat{m}_i^2$ with the thermally corrected mass $\hat{M}_i^2\equiv \hat{m}_i^2+\Pi_i T^2$ throughout the entire effective potential, with $\Pi_i$ for the relevant species in the plasma given by\footnote{Technically what is described here is the Parwani scheme~\cite{Parwani:1991gq}, which is adopted in our numerical calculations.
An alternate prescription is the Arnold-Espinosa resummation scheme~\cite{Arnold:1992rz}, which amounts to add another term $V_{\rm ring} \equiv - T \sum_i n_i(\hat{M}_i^3 - \hat{m}_i^3)/(12\pi)$ with the summation runs over the bosonic degrees of freedom.
% This is equivalent to replacing $\hat{m}_B^2 \rightarrow \hat{M}_B^2$ only in the finite-temperature piece of the effective potential when cons.
In doing so, only the problematic (soft) zero modes are resummed.
In contrast, the Parwani scheme resums both soft and hard modes by using the thermally corrected mass throughout the effective potential. 
So long as one is always within the EFT regime of validity, using either scheme should yield essentially the same results, with any differences appearing at higher order.
For recent progress on resummations, see Ref.~\cite{Bahl:2024ykv,Bittar:2025lcr}}:
\begin{equation}
\begin{split}
    & \Pi_{\rm GB}^L = \frac{11}{6} \text{diag} ( g^2, g^2, g^2, {g'}^{\, 2} ) \,,\\
    & \mathbb{Z}_2\text{ singlet scalar model:} \\
    & \Pi_S = \frac{\lambda_S}{4} + \frac{\lambda_{SH}}{6} \,, \\
    & \Pi_h = \Pi_{\chi_{1,2,3}} = \frac{\lambda_H}{2} + \frac{\lambda_{SH}}{24} + \frac{{g'}^{\,2} + 3 g^2}{16} + \frac{y_t^2}{4} \,, \\
    & \text{DarkCPV model:} \\
    & \Pi_s = \Pi_a = \frac{\lambda}{12} + \frac{\lambda_\SC}{3} + \frac{\lambda_{\SC H}}{6} \\
    & \Pi_h = \Pi_{\chi_{1,2,3}} = \frac{3 g^2 + g'^2}{16} + \frac{y_t^2}{4} + \frac{\lambda_H}{2} + \frac{\lambda_{\SC H} }{12}
\end{split}
\end{equation}
where $\Pi_{\rm GB}^L$ is added to the gauge boson masses (squared) in the gauge basis. Note that fermions are protected from large thermal masses by chiral symmetry while the transverse modes of gauge bosons are protected by gauge symmetry, so $\Pi_t \simeq \Pi_{W}^T \simeq \Pi_Z^T \simeq 0$ to leading order.
\section{Semi-classical Frameworks of EWBG Calculation}
\label{sec:app_bau}

We start from the Boltzmann equation Eq.~\eqref{eq:BAU Boltzmann}.
Substituting the general ansatz Eq.~\eqref{eq:ansatz general} into the Boltzmann equation, we get,
\begin{equation}
    v_g\left[\left(\gamma_w(\partial_z E_w)_{k_z}-\partial_z\mu\right)f_i^\prime + \partial_z\delta f\right] + F_{z}\left[\gamma_w \left((\partial_{k_z}E_w)_z+v_w\right)f_i^\prime + \partial_{k_z}\delta f\right]=C[f_i, f_j,...],
    \label{eq:be1}
\end{equation}
where $f_{i}^\prime\equiv\partial f_i/\partial(\gamma_w E_w)$. We observe the cancellation of the following terms due to energy conservation in the wall frame:
\begin{equation}
    v_g (\partial_z E_w)_{k_z} + F_{z}(\partial_{k_z}E_w)_z=\Dot{z} (\partial_z E_w)_{k_z} + \Dot{k_z}(\partial_{k_z}E_w)_z=dE_w/dt=0.
\end{equation}
Therefore, Eq.~(\ref{eq:be1}) is simplified to,
\begin{equation}
    f_i^\prime\left(v_w \gamma_w F_{z}-v_g\partial_z\mu\right)+v_g\,\partial_z\delta f + F_{z}\,\partial_{k_z}\delta f =C[f_i, f_j,...].
    \label{eq:be_precise}
\end{equation}
It is still difficult to solve this equation directly, since $f_i^\prime$ is a non-linear function of the unknowns $\mu$ and $\delta f$. To proceed, we must linearize this equation, upon which the calculations of \cite{Cline:2000nw, Cline:2020jre} diverge.

Moreover, $\mu$ and $\delta f$ can be decomposed into the CP-even and CP-odd components as, 
\begin{align}
    \mu=\mu_e + s_{k_0}\mu_o,\quad \delta f = \delta f_e + s_{k_0}\delta f_o\,,\label{eq:mu_df_decomp}\end{align}
where $s_{k_0}=+1\,(-1)$ for particle (anti-particle). According to \cref{eq:Fz_E0}, $F_z$ can also be decomposed into $F_z=F_e + s_{k_0} F_o$, with
    \begin{align}
    F_e=-\frac{(|m|^2)'}{2 E_0},\quad F_o=s\left[\frac{(|m|^2\theta')'}{2 E_0 E_z} - \frac{|m|^2(|m|^2)'\theta'}{4 E_0^3 E_z}\right].
    \label{eq:Fz_decomp}
\end{align}
The Boltzmann equations for the CP-even and CP-odd components decouple once \cref{eq:be_precise} is linearized. 

\subsection{L.O. Approximation}

The leading order (L.O.) approximation~\cite{Cline:2000nw,Fromme:2006wx} assumes a small $v_w$ and simplifies the Lorentz boost in \cref{eq:ansatz general} to a Galilean transformation, $v_g \to v_g + v_w$, $\gamma_w\approx 1$. Transforming from the rest frame of the wall to the plasma frame, and keeping the leading order terms of $\mu$, $\delta f$ and the force $F_z$, \cref{eq:be_precise} simplifies to,
\begin{equation}
   f_0^\prime v_w F_{z} +\left(v_g+v_w\right)\left(-f_0^\prime\partial_z\mu + \partial_z\delta f\right)=C[f_i, f_j, ...],
    \label{eq:BE_cjk}
\end{equation}
where $f_0$ is the Fermi-Dirac/Bose-Einstein distribution $f_0=(e^{\beta E_0}\pm 1)^{-1}$, and $f_0'=df_0/dE_0$.
Note that $E_w$ has been replaced by $E_0$, ignoring the energy correction $\Delta E$ due to the CP-violating effects, which does not contribute at the leading order. The linearized Boltzmann equation can be decomposed into the CP-even and CP-odd equations as discussed above. Since the CP-even and CP-odd equations take the same form, they can be treated in the same way as discussed below, dropping the $e$ and $o$ subscripts.

Integrate both sides of \cref{eq:BE_cjk} over momentum, weighting by $1$ and $k_z/E_0$ respectively, and substituting $v_g=k_z/E_0$, we get two independent moment equations:
\begin{align}
    -v_w\partial_z\mu +\langle\frac{k_z}{E_0}\partial_z\delta f\rangle &= \langle C\rangle,\label{eq:moment0}\\
    -\langle\frac{k_z^2}{E_0^2}f_0^\prime\rangle\partial_z\mu
    +v_w\langle\frac{k_z}{E_0}\partial_z\delta f\rangle + v_w\langle\frac{k_z}{E_0}F_{z}f_0^\prime\rangle&=\langle\frac{k_z}{E_0}C\rangle,
    \label{eq:moment}
\end{align}
where
\begin{equation}
    \langle X\rangle\equiv\frac{\int\! d^3 k\, X}{\int\! d^3 k\, f_0^\prime}.
\end{equation}
In deriving \cref{eq:moment0} and \cref{eq:moment}, we use the fact that some terms are vanishing: $\int\!d^3 k\, v_g f_0^\prime=\int\!d^3 k\,(\partial_{k_z}E_0)_z f_0^\prime=\int\!d^3 k\, df_0/dk_z=0$, $\int\!d^3 k\,\delta f^\prime=(\int\!d^3k\,\delta f)^\prime=0$, $\int\!d^3k\,f_0^\prime =0$ since $f_0$ is symmetric in $\pm k_z$. A further simplification can be made by linearizing the collision terms as:
\begin{align}
    &\langle \frac{k_z}{E_0}C_i\rangle\simeq -\Gamma_i^{\rm tot}u;\\
    &\langle C_i\rangle\simeq K_0\sum_c \Gamma_{i}^c\sum_{j}s_{j}^c\frac{\mu_j^c}{T};\\
    &K_0\equiv-\frac{\int d^3 k f_0}{\int d^3 k f_0^\prime},
\end{align}
where we introduced a new variable $u\equiv \langle(k_z/E_0)\delta f\rangle$.
To the leading order of the WKB approximation, we can take $\langle(k_z/E_0)\partial_z\delta f\rangle\simeq \partial_z u$. \cref{eq:moment0} and \cref{eq:moment} can then be written in the matrix form of Eq.~\eqref{eq:transport}
with $\omega=(\mu,u)^T$, with the coefficients given by Eq.~\eqref{eq:coefficient CJK}.
Such a linear system has a singularity at finite $v_w$ as
\begin{align}
    \det A \to 0 ~\text{ at }~v_w \to \sqrt{\langle\frac{k_z^2}{E_0^2}f_0^\prime\rangle}.
\end{align}

\subsection{Full Treatment}

To extend the calculation to higher bubble wall velocities, Ref.~\cite{Cline:2020jre} reinstated the Lorentz transformation between the wall frame and the plasma frame for the zeroth order equilibrium distribution,
\begin{align}
    f_{0w}=\frac{1}{e^{\beta\gamma_w(E_0+v_w k_z)}\pm 1} \,.
\end{align}
With the decomposition of \cref{eq:mu_df_decomp} and \cref{eq:Fz_decomp}, the departure of the general ansatz \cref{eq:ansatz general} from $f_{0w}$ can also be decomposed into the CP-even and CP-odd contributions as
\begin{align}
    f=f_{0w}+\Delta f_e + s_{k_0}\Delta f_o,
    \label{eq:f_ansatz_expand}
\end{align}
with $\Delta f_e$ and $\Delta f_o$ expanded up to the second order in spatial derivatives as
\begin{align}
&\Delta f_e \approx -\mu_e f_{0w}' + \delta f_e,\\
&\Delta f_o \approx (-\mu_o + s\gamma_w\Delta E) f_{0w}' - sf_{0w}''\gamma_w\Delta E \mu_e + \delta f_o.
\end{align}
When counting the spatial derivative orders, the CP-even terms ($\mu_e$, $\delta f_e$) are counted as first order, as they are sourced from the first-order CP-even force. 
The CP-odd terms ($\mu_o$, $\delta f_o$, $\Delta E$) are counted as second order, as they are sourced from the second-order CP-odd force.

Substitute \cref{eq:f_ansatz_expand} into \cref{eq:be_precise}, we can linearize the Boltzmann equation and decouple the CP-even and CP-odd equations, which are formulated as
\begin{align}
    L[\mu_i,\delta f_i] = \mathcal{S}_i + \delta \mathcal{C}_i,~~~~i=e,o.
\end{align}
The Liouville operator takes the same form for the CP-even and CP-odd equations as
\begin{align}
    L[\mu,\delta f]\equiv -\frac{k_z}{E_0}f_{0w}'\partial_z\mu + v_w\gamma_w\frac{(|m|^2)'}{2E_0}f_{0w}''\mu + \frac{k_z}{E_0}\partial_z \delta f - \frac{(|m|^2)'}{2E_0}\partial_{k_z}\delta f.
\end{align}
The source terms take the following forms:
\begin{align}
    &\mathcal{S}_e=v_w\gamma_w\frac{(|m|^2)'}{2E_0}f_{0w}',\label{eq:S_cpeven}\\
    &\mathcal{S}_o=-v_w\gamma_w s\frac{(|m|^2\theta')'}{2E_0 E_z}f_{0w}' + v_w \gamma_w s \frac{|m|^2(|m|^2)'\theta'}{4 E_0^2 E_z}\left(\frac{f_{0w}'}{E_0}-\gamma_w f_{0w}''\right).\label{eq:S_cpodd}
\end{align}
In deriving $\mathcal{S}_o$, some terms proportional to $\mu_e$ and $\delta f_e$ are dropped in Ref.~\cite{Cline:2020jre}. The CP-even and CP-odd equations can again be treated in the same way as discussed below, dropping the $e,o$ subscripts.

Integrating both sides of the Boltzmann equation over momentum, weighting by 1 and $k_z/E_0$ respectively, we get the following moment equations:
\begin{align}
    -\langle \frac{k_z}{E_0}f_{0w}^\prime\rangle\mu^\prime + \gamma_w v_w(|m|^2)^\prime\langle\frac{f_{0w}^{\prime\prime}}{2E_0}\rangle\mu+u_1^\prime&=\langle\mathcal{S}\rangle +\langle \delta\mathcal{C}[f]\rangle,
    \label{eq:be_moment1}\\
    -\langle \frac{k_z^2}{E_0^2}f_{0w}^\prime\rangle\mu^\prime + \gamma_w v_w(|m|^2)^\prime\langle\frac{k_z f_{0w}^{\prime\prime}}{2E_0^2}\rangle\mu + u_2^\prime + (|m|^2)^\prime\langle\frac{1}{2E_0^2}\delta f\rangle& =\langle\frac{k_z}{E_0}\mathcal{S}\rangle + \langle\frac{k_z}{E_0}\delta\mathcal{C}[f]\rangle,
    \label{eq:be_moment2}
\end{align}
with the new variable $u_\ell$ defined as
\begin{equation}
    u_\ell\equiv\langle\left(\frac{k_z}{E_0}\right)^\ell\delta f\rangle.
\end{equation}
The last term on the LHS of Eq.~(\ref{eq:be_moment2}) cannot be directly evaluated since we do not know the momentum dependence of $\delta f$. We factorize it as
\begin{equation}
    \langle\frac{1}{2E_0^2}\delta f\rangle=\Bar{R}u_1,
\end{equation}
with $u_2=-v_w u_1$. \cref{eq:be_moment1} and \cref{eq:be_moment2} define the coefficient matrices given in \cref{eq:coefficient CK20}.

Finally, we comment that in this paper, we have weighted the Boltzmann equation by $1$ and $k_z/E_0$.
Recently, Ref.~\cite{Kainulainen:2024qpm} has considered higher orders of weighting over momentum.
We leave this improvement for future work.

\bibliographystyle{JHEP}
\bibliography{main.bib}
\end{document}